# Hubbard Model with Inter-Site Kinetic Correlations

Grzegorz Górski  and  Jerzy Mizia

Institute of Physics, Rzeszów University, ul. Rejtana 16A, 35-959 Rzeszów, Poland

We introduced the inter-site electron-electron correlation to the Hubbard III approximation. This correlation was excluded in the Hubbard III approximation and also in the equivalent coherent potential approximation. Including it brings two spin dependent effects: the bandwidth correction and the bandshift correction, which both stimulate the ferromagnetic ground state. The bandshift correction factor causes an exchange splitting between the spin-up and spin-down spectrum, and its role is similar to the exchange interaction in the classic Stoner model. The spin dependent bandwidth correction lowers the kinetic energy of electrons by decreasing the majority spin bandwidth for some electron occupations with respect to the minority spin bandwidth. In certain conditions it can lead to ferromagnetic alignment. A gain in the kinetic energy achieved in this way is the opposite extreme to the effect of a gain in potential energy due to exchange splitting. The bandshift factor is a dominant force behind the ferromagnetism. The influence of the bandwidth factor is too weak to create ferromagnetism and the only result is the correction to the classic coherent potential approximation in favor of ferromagnetism.

## I. INTRODUCTION

The Hubbard model[1-3] is extensively used to analyze strong electron correlations in the narrow energy bands. Of special interest are applications of this model to itinerant ferromagnetism, metal-insulator phase transitions, or high-$T_C$ superconductivity.

This model describes the dynamics of electrons in crystals. In its simple original version it uses such quantities as the electron dispersion energy $\varepsilon_k$, and the on-site Coulomb interaction $U = (i,i|1/r|i,i)$, where $i$ is the lattice site index. The electron dispersion energy is the Fourier transform of the inter-site hopping integral $t_{ij}$. In the systems with strong correlation the on-site Coulomb interaction $U$ causes a split of the spin band into two sub-bands: lower sub-band centered around the atomic level $T_0$ (we assume $T_0 = 0$), and the upper sub-band centered around the level $T_0 + U$.

Despite its simplicity there is no exact solution to the Hubbard Hamiltonian with the exception of the one dimensional system,[4] therefore for many years a variety of different approximations have been used for this model. The important one was the Hubbard I approximation,[1] which is rigorous in the atomic ($t_{ij} = 0$) limit and in the band ($U = 0$) limit. Unfortunately this approximation produces a band split into two sub-bands separated by an energy gap, even for arbitrarily small Coulomb repulsion. The additional odd feature of this approximation is an infinite lifetime of the pseudo-particles caused by the real value of the self-energy. These two negative features are the result of the assumption that the dominant correlation takes place only between two electrons on the same lattice site. In the Green function language it is implemented by assuming that the Green function involving more than two atomic sites can be approximated by the single site average multiplied by the two sites Green function. Thus in Eq. (7) below the higher order Green function $\langle\langle \hat{n}_{i-\sigma} c_{l\sigma} ; c_{j\sigma}^+ \rangle\rangle_\varepsilon$ is approximated by $\langle \hat{n}_{i-\sigma} \rangle \langle\langle c_{l\sigma} ; c_{j\sigma}^+ \rangle\rangle_\varepsilon$.

Further attempts to improve the solution or simply to obtain a ferromagnetic ground state included the mean field approximation, the so-called Hubbard III approximation[3] or equivalent coherent potential approximation (CPA),[5,6] the slave-boson method,[7,8] and e.g. dynamical mean-field theory.[9] The Hubbard model was also analyzed directly by the numerical Quantum Monte Carlo simulation.[10-12] All these attempts did not bring the desired solution. Such a solution would have a ferromagnetic ground state obtained under credible approximations.

To describe realistic systems where physical phenomena like ferromagnetism or high temperature superconductivity exist the inter-site Coulomb interactions were added to the simple Hubbard model.[13-17] These are: the charge-charge interaction $V=(i,j|1/r|i,j)$, inter-site exchange interaction $J=(i,j|1/r|j,i)$ and the hopping interaction $\Delta t=(i,i|1/r|j,i)$. For high temperature superconducting compounds, for which there is more than one band involved in a given physical property, two-band,[18-21] three-band,[22-27] etc. versions of the Hubbard model were used. In the models describing the copper-oxygen $CuO_2$ plane, the hopping integrals between different orbitals were also introduced, e.g. copper-oxygen hopping integral in the $CuO_2$ plane. All these additional interactions or different hopping integrals will not be considered in this paper, but we will focus on the simple basic Hubbard model with interaction $U$ and hopping integral $t_{ij}$.

As mentioned above, Hubbard introduced the approximation called the Hubbard III approximation.[3] This approximation did not produce the ferromagnetic ground state (see Refs. 28 and 29). The best way to prove it is to translate the Hubbard III result to the CPA approximation,[6] and to analyze the solution in this language.[28]

In this paper we will describe in great detail the Hubbard III approximation with the included inter-site kinetic correlation functions $\langle c^+_{i-\sigma} c_{j-\sigma} \rangle$ and $\langle \hat{n}_{i\sigma} c^+_{i-\sigma} c_{j-\sigma} \rangle$. These correlations were originally ignored in the Hubbard approach and in most of the subsequent papers devoted to this model. They were considered by Roth[30] and Nolting and co-workers[31-35] within the framework of the two pole approximation, which eventually led to the spectral density approach (SDA)[31-33] and the modified alloy analogy (MAA).[34,35] The basis of the SDA method is the Roth's two-pole approximation,[30] which gives the two-pole ansatz for the single-particle spectral density function. The SDA approximation brings the magnetic results in the strong-coupling limit, but it neglects the quasiparticle damping. It is the extension of the Hubbard I approximation. Such an approach perhaps can be justified for the systems with strong correlation. To correct the SDA approximation Nolting and co-workers proposed the MAA method,[34,35] which is a combination of the SDA and CPA methods. In the CPA method there are two spin independent atomic levels $V_1=T_0$ and $V_2=T_0+U$, which in the MAA method are replaced by two atomic levels dependent on occupation and spin. We will analyze results of the MAA method and show that they can be obtained as a simplified version of our model in which the inter-site correlations are included directly into the Hubbard III or CPA scheme.

The paper is organized as follows. In Sec. II the general Green function chain equations for the Hubbard model are recalled. In Sec. III and Sec. IV the solution of this chain within the framework of the Hubbard III approximation with included inter-site correlation is given. In Sec. V the bandwidth and bandshift corrections are calculated using the Hartree-Fock (H-F) approximation and the more rigorous high approximation. Based on these results the magnetic analysis of the system ground state is performed in Sec. VI, showing the possibility of ferromagnetic transition at some electron occupations under the on-site interaction $U$ alone. Discussion and the comparison with the results of SDA and MAA methods are presented in Sec. VII.

## II. GREEN FUNCTION CHAIN EQUATIONS FOR THE HUBBARD MODEL

The simple Hubbard Hamiltonian in the real space representation has the following form[3]

$$H = -\sum_{ij\sigma} t_{ij} c_{i\sigma}^+ c_{j\sigma} + \frac{U}{2} \sum_{i\sigma} \hat{n}_{i\sigma} \hat{n}_{i-\sigma} - \mu \sum_{i\sigma} \hat{n}_{i\sigma} \quad , \tag{1}$$

where $t_{ij}$ – hopping integral between the $i$-th and $j$-th lattice site. The operator $c_{i\sigma}^+ (c_{i\sigma})$ is creating (annihilating) an electron with spin $\sigma$ on the $i$-th lattice site, $\hat{n}_{i\sigma} = c_{i\sigma}^+ c_{i\sigma}$ is the electron number operator for electrons with spin $\sigma$ on the $i$-th lattice site, and $\mu$ is the chemical potential. The term with chemical potential will be ignored below since it will appear in the Fermi-Dirac statistics.

The equation of motion for the Green function has the following form[1,36]

$$\varepsilon \langle\langle A;B \rangle\rangle_\varepsilon = \langle [A,B]_+ \rangle + \langle\langle [A,H]_-;B \rangle\rangle_\varepsilon \quad , \tag{2}$$

where $A$ and $B$ are the single operators or their products.

Using in relation (2) the Hamiltonian (1) we obtain the following equation of motion for the Green function $\langle\langle c_{i\sigma}; c_{j\sigma}^+ \rangle\rangle_\varepsilon$:

$$\varepsilon \langle\langle c_{i\sigma}; c_{j\sigma}^+ \rangle\rangle_\varepsilon = \delta_{ij} - \sum_l t_{il} \langle\langle c_{l\sigma}; c_{j\sigma}^+ \rangle\rangle_\varepsilon + U \langle\langle \hat{n}_{i-\sigma} c_{i\sigma}; c_{j\sigma}^+ \rangle\rangle_\varepsilon \quad . \tag{3}$$

For the higher order Green function $\langle\langle \hat{n}_{i-\sigma} c_{i\sigma}; c_{j\sigma}^+ \rangle\rangle_\varepsilon$ appearing above on the right hand side, the Hubbard III approximation will be used together with the Hubbard notation for the electron number operators[3]

$$\hat{n}_{i\sigma}^+ \equiv \hat{n}_{i\sigma} \quad , \qquad \hat{n}_{i\sigma}^- \equiv 1 - \hat{n}_{i\sigma} \quad , \qquad \sum_{\alpha=\pm} \hat{n}_{i\sigma}^\alpha = 1 \quad , \tag{4}$$

and for the two resonant energies

$$\varepsilon_+ \equiv U, \qquad \varepsilon_- \equiv 0 \ . \tag{5}$$

The same notation as in Eq. (4) will be introduced for the average electron occupations

$$n_\sigma^+ = \langle \hat{n}_{i\sigma}^+ \rangle \equiv n_\sigma \ , \qquad n_\sigma^- = \langle \hat{n}_{i\sigma}^- \rangle \equiv 1 - n_\sigma \ . \tag{6}$$

Applying equation (2) to the higher order Green function: $\langle\langle \hat{n}_{i-\sigma}^\alpha c_{i\sigma}; c_{j\sigma}^+ \rangle\rangle_\varepsilon$ ($\alpha = \pm$), with notation of Eqs. (4), (5) and (6), we obtain the equation[3]

$$\begin{aligned}
\varepsilon \langle\langle \hat{n}_{i-\sigma}^\alpha c_{i\sigma}; c_{j\sigma}^+ \rangle\rangle_\varepsilon = & n_{-\sigma}^\alpha \left( \delta_{ij} - \sum_l t_{il} \langle\langle c_{l\sigma}; c_{j\sigma}^+ \rangle\rangle_\varepsilon \right) + \varepsilon_\alpha \langle\langle \hat{n}_{i-\sigma}^\alpha c_{i\sigma}; c_{j\sigma}^+ \rangle\rangle_\varepsilon \\
& - \sum_l t_{il} \langle\langle (\hat{n}_{i-\sigma}^\alpha - n_{-\sigma}^\alpha) c_{l\sigma}; c_{j\sigma}^+ \rangle\rangle_\varepsilon \\
& - \xi_\alpha \sum_l t_{il} \left( \langle\langle c_{i-\sigma}^+ c_{l-\sigma} c_{i\sigma}; c_{j\sigma}^+ \rangle\rangle_\varepsilon - \langle\langle c_{l-\sigma}^+ c_{i-\sigma} c_{i\sigma}; c_{j\sigma}^+ \rangle\rangle_\varepsilon \right)
\end{aligned} \tag{7}$$

where $\xi_\pm = \pm 1$.

### III. HUBBARD III APPROXIMATION WITH INTER-SITE KINETIC CORRELATION

Taking into account only the first two terms on the right hand side of Eq. (7) gives the Hubbard I approximation.[1] Including the third term in Eq. (7), which comes from the commutator $[c_{i\sigma}, H]_-$ in the equation of motion, leads to what is known as the "scattering effect". The last term, which comes from the commutator $[\hat{n}_{i-\sigma}, H]_-$ in the equation of motion, gives the "resonance broadening effect".

In further considerations of the scattering effect and the resonance broadening effect the new averages of the kinetic type $\langle c_{i-\sigma}^+ c_{l-\sigma} \rangle$ and $\langle \hat{n}_{i\sigma} c_{i-\sigma}^+ c_{j-\sigma} \rangle$ will be kept. This will result in corrections to the Hubbard scattering and resonance broadening effects, which are the function of these averages.

**The scattering effect**

To consider this effect we ignore the last term in Eq. (7) and search for the solution of function $\langle\langle (\hat{n}_{i-\sigma}^\alpha - n_{-\sigma}^\alpha) c_{l\sigma}; c_{j\sigma}^+ \rangle\rangle_\varepsilon$. Using Eq. (4) one can write this Green function as

$$\langle\langle (\hat{n}_{i-\sigma}^\alpha - n_{-\sigma}^\alpha) c_{l\sigma}; c_{j\sigma}^+ \rangle\rangle_\varepsilon = \sum_{\beta=\pm} \langle\langle (\hat{n}_{i-\sigma}^\alpha - n_{-\sigma}^\alpha) \hat{n}_{l-\sigma}^\beta c_{l\sigma}; c_{j\sigma}^+ \rangle\rangle_\varepsilon \ . \tag{8}$$

The equation of motion for the Green function $\langle\langle(\hat{n}^{\alpha}_{i-\sigma} - n^{\alpha}_{-\sigma})\hat{n}^{\beta}_{l-\sigma}c_{l\sigma}; c^{+}_{j\sigma}\rangle\rangle_{\varepsilon}$ has the following form

$$\varepsilon\langle\langle(\hat{n}^{\alpha}_{i-\sigma} - n^{\alpha}_{-\sigma})\hat{n}^{\beta}_{l-\sigma}c_{l\sigma}; c^{+}_{j\sigma}\rangle\rangle_{\varepsilon} = \delta_{jl}\langle(\hat{n}^{\alpha}_{i-\sigma} - n^{\alpha}_{-\sigma})\hat{n}^{\beta}_{l-\sigma}\rangle$$
$$+\varepsilon_{\beta}\langle\langle(\hat{n}^{\alpha}_{i-\sigma} - n^{\alpha}_{-\sigma})\hat{n}^{\beta}_{l-\sigma}c_{l\sigma}; c^{+}_{j\sigma}\rangle\rangle_{\varepsilon} - \sum_{m}t_{ml}\langle\langle(\hat{n}^{\alpha}_{i-\sigma} - n^{\alpha}_{-\sigma})\hat{n}^{\beta}_{l-\sigma}c_{m\sigma}; c^{+}_{j\sigma}\rangle\rangle_{\varepsilon} \quad . \quad (9)$$

+other terms

In the original Hubbard model[3] the first term on the right hand side of Eq. (9) was assumed to be zero. In our model this average will be written as

$$\langle(\hat{n}^{\alpha}_{i-\sigma} - n^{\alpha}_{-\sigma})\hat{n}^{\beta}_{l-\sigma}\rangle = \xi_{\alpha}\xi_{\beta}(\langle\hat{n}_{i-\sigma}\hat{n}_{l-\sigma}\rangle - n^{2}_{-\sigma}) = \xi_{\alpha}\xi_{\beta}B^{\sigma}_{il} \quad . \quad (10)$$

It will be evaluated later on in the H-F approximation, Eq. (64), and in the high approximation, Eq. (81). In both these approximations $B^{\sigma}_{il}$ is expressed by the inter-site averages of the type $\langle c^{+}_{i-\sigma}c_{l-\sigma}\rangle$ and $\langle \hat{n}_{i\sigma}c^{+}_{i-\sigma}c_{l-\sigma}\rangle$, which are assumed to be nonzero and will be evaluated below.

For the function $\langle\langle(\hat{n}^{\alpha}_{i-\sigma} - n^{\alpha}_{-\sigma})\hat{n}^{\beta}_{l-\sigma}c_{m\sigma}; c^{+}_{j\sigma}\rangle\rangle_{\varepsilon}$ appearing in Eq. (9), the following approximation will be used

$$\langle\langle(\hat{n}^{\alpha}_{i-\sigma} - n^{\alpha}_{-\sigma})\hat{n}^{\beta}_{l-\sigma}c_{m\sigma}; c^{+}_{j\sigma}\rangle\rangle_{\varepsilon} \cong n^{\beta}_{-\sigma}\langle\langle(\hat{n}^{\alpha}_{i-\sigma} - n^{\alpha}_{-\sigma})c_{m\sigma}; c^{+}_{j\sigma}\rangle\rangle_{\varepsilon}$$
$$+\langle(\hat{n}^{\alpha}_{i-\sigma} - n^{\alpha}_{-\sigma})\hat{n}^{\beta}_{l-\sigma}\rangle\langle\langle c_{m\sigma}; c^{+}_{j\sigma}\rangle\rangle_{\varepsilon} \quad , \quad (11)$$

where the first term on the right hand side is the original Hubbard III term. The additional second term is the result of assuming that $B^{\sigma}_{il} \neq 0$.

Inserting to Eq. (9) the approximations (10) and (11) one obtains the relation

$$(\varepsilon - \varepsilon_{\beta})\langle\langle(\hat{n}^{\alpha}_{i-\sigma} - n^{\alpha}_{-\sigma})\hat{n}^{\beta}_{l-\sigma}c_{l\sigma}; c^{+}_{j\sigma}\rangle\rangle_{\varepsilon} = \xi_{\alpha}\xi_{\beta}B^{\sigma}_{il}\left(\delta_{jl} - \sum_{m}t_{ml}\langle\langle c_{m\sigma}; c^{+}_{j\sigma}\rangle\rangle_{\varepsilon}\right)$$
$$-\sum_{m}t_{ml}n^{\beta}_{-\sigma}\langle\langle(\hat{n}^{\alpha}_{i-\sigma} - n^{\alpha}_{-\sigma})c_{m\sigma}; c^{+}_{j\sigma}\rangle\rangle_{\varepsilon} \quad . \quad (12)$$

Dividing both sides of this equation by $(\varepsilon - \varepsilon_{\beta})$ and summing up over $\beta = \pm$ one has

$$\langle\langle(\hat{n}^{\alpha}_{i-\sigma} - n^{\alpha}_{-\sigma})c_{l\sigma}; c^{+}_{j\sigma}\rangle\rangle_{\varepsilon} = \xi_{\alpha}C_{-}(\varepsilon)B^{\sigma}_{il}\left(\delta_{jl} - \sum_{m}t_{ml}\langle\langle c_{m\sigma}; c^{+}_{j\sigma}\rangle\rangle_{\varepsilon}\right)$$
$$-\frac{1}{F^{\sigma}_{H,0}(\varepsilon)}t_{il}\langle\langle(\hat{n}^{\alpha}_{i-\sigma} - n^{\alpha}_{-\sigma})c_{i\sigma}; c^{+}_{j\sigma}\rangle\rangle_{\varepsilon} - \frac{1}{F^{\sigma}_{H,0}(\varepsilon)}\sum_{m\neq i}t_{ml}\langle\langle(\hat{n}^{\alpha}_{i-\sigma} - n^{\alpha}_{-\sigma})c_{m\sigma}; c^{+}_{j\sigma}\rangle\rangle_{\varepsilon} \quad ,(13)$$

where

$$\frac{1}{F^{\sigma}_{H,0}(\varepsilon)} = \frac{n^{+}_{-\sigma}}{\varepsilon - \varepsilon_{+}} + \frac{n^{-}_{-\sigma}}{\varepsilon - \varepsilon_{-}} \quad , \quad (14)$$

$$C_-(\varepsilon) = \frac{1}{\varepsilon - \varepsilon_+} - \frac{1}{\varepsilon - \varepsilon_-} \quad . \tag{15}$$

Following the Hubbard I approximation one can write that

$$\delta_{jl} - \sum_m t_{ml} \langle\langle c_{m\sigma}; c_{j\sigma}^+ \rangle\rangle_\varepsilon = F_{H,0}^\sigma(\varepsilon) \langle\langle c_{l\sigma}; c_{j\sigma}^+ \rangle\rangle_\varepsilon \quad . \tag{16}$$

Inserting Eq. (16) to Eq. (13) one obtains the following relation

$$\langle\langle (\hat{n}_{i-\sigma}^\alpha - n_{-\sigma}^\alpha) c_{l\sigma}; c_{j\sigma}^+ \rangle\rangle_\varepsilon = \xi_\alpha C_-(\varepsilon) F_{H,0}^\sigma(\varepsilon) B_{il}^\sigma \langle\langle c_{l\sigma}; c_{j\sigma}^+ \rangle\rangle_\varepsilon$$
$$- \frac{1}{F_{H,0}^\sigma(\varepsilon)} \left[ t_{il} \langle\langle (\hat{n}_{i-\sigma}^\alpha - n_{-\sigma}^\alpha) c_{i\sigma}; c_{j\sigma}^+ \rangle\rangle_\varepsilon + \sum_{m \neq i} t_{ml} \langle\langle (\hat{n}_{i-\sigma}^\alpha - n_{-\sigma}^\alpha) c_{m\sigma}; c_{j\sigma}^+ \rangle\rangle_\varepsilon \right] \quad . \tag{17}$$

Equation (17) is analogous to Eq. (25) of Hubbard,[3] but it contains additionally the bandwidth correction $B_{il}^\sigma$. Solution of Eq. (17) is the Hubbard solution, with the additional first term on the right responsible for the bandwidth correction. Hence the Green function $\langle\langle (\hat{n}_{i-\sigma}^\alpha - n_{-\sigma}^\alpha) c_{l\sigma}; c_{j\sigma}^+ \rangle\rangle_\varepsilon$ will have the form

$$\langle\langle (\hat{n}_{i-\sigma}^\alpha - n_{-\sigma}^\alpha) c_{l\sigma}; c_{j\sigma}^+ \rangle\rangle_\varepsilon = -\sum_m W_{lm,i}^\sigma(\varepsilon) t_{mi} \langle\langle (\hat{n}_{i-\sigma}^\alpha - n_{-\sigma}^\alpha) c_{i\sigma}; c_{j\sigma}^+ \rangle\rangle_\varepsilon$$
$$+ \xi_\alpha C_-(\varepsilon) F_{H,0}^\sigma(\varepsilon) B_{il}^\sigma \langle\langle c_{l\sigma}; c_{j\sigma}^+ \rangle\rangle_\varepsilon \quad , \tag{18}$$

where

$$W_{lm,i}^\sigma(\varepsilon) = g_{lm}^\sigma(\varepsilon) - \frac{g_{li}^\sigma(\varepsilon) g_{im}^\sigma(\varepsilon)}{g_{ii}^\sigma(\varepsilon)} \quad , \tag{19}$$

and

$$g_{ij}^\sigma(\varepsilon) = \sum_k \frac{\exp[i\mathbf{k} \cdot (\mathbf{r}_i - \mathbf{r}_j)]}{F_{H,0}^\sigma(\varepsilon) - \varepsilon_k} \quad . \tag{20}$$

Equation (18) differs from the Hubbard's expression for scattering effect (Eq. (26) in Ref. 3) by the second term on the right hand side, which includes the $B_{il}^\sigma$ factor.

Inserting the Green function from Eq. (18) to Eq. (7) (we are still ignoring the last term in Eq. (7), which will be dealt with in the next section on resonance broadening effect) one obtains the following equation for the Green function $\langle\langle \hat{n}_{i-\sigma}^\alpha c_{i\sigma}; c_{j\sigma}^+ \rangle\rangle_\varepsilon$

$$(\varepsilon - \varepsilon_\alpha) \langle\langle \hat{n}_{i-\sigma}^\alpha c_{i\sigma}; c_{j\sigma}^+ \rangle\rangle_\varepsilon + \xi_\alpha \Omega_\sigma(\varepsilon) \left( n_{-\sigma}^+ \langle\langle \hat{n}_{i-\sigma}^- c_{i\sigma}; c_{j\sigma}^+ \rangle\rangle_\varepsilon - n_{-\sigma}^- \langle\langle \hat{n}_{i-\sigma}^+ c_{i\sigma}; c_{j\sigma}^+ \rangle\rangle_\varepsilon \right)$$
$$= n_{-\sigma}^\alpha \left( \delta_{ij} - \sum_l t_{il} \langle\langle c_{l\sigma}; c_{j\sigma}^+ \rangle\rangle_\varepsilon \right) + \xi_\alpha \sum_l B_{il,\sigma}^S(\varepsilon) \langle\langle c_{l\sigma}; c_{j\sigma}^+ \rangle\rangle_\varepsilon \quad , \tag{21}$$

where

$$\Omega_\sigma(\varepsilon) = \sum_{l,m} t_{il} W^\sigma_{lm,i}(\varepsilon) t_{mi} \; , \tag{22}$$

$$B^S_{il,\sigma}(\varepsilon) = C_-(\varepsilon) F^\sigma_{H,0}(\varepsilon)(-t_{il})(\langle \hat{n}_{i-\sigma} \hat{n}_{l-\sigma} \rangle - n^2_{-\sigma}) \; . \tag{23}$$

**The resonance broadening effect**

The resonance broadening effect is described by the Green function $\langle\langle c^\pm_{l-\sigma} c^\mp_{i-\sigma} c_{i\sigma}; c^+_{j\sigma} \rangle\rangle_\varepsilon$ appearing in the last term of Eq. (7). Using Eq. (4) we can write that

$$\langle\langle c^\pm_{l-\sigma} c^\mp_{i-\sigma} c_{i\sigma}; c^+_{j\sigma} \rangle\rangle_\varepsilon = \sum_{\alpha=\pm} \langle\langle \hat{n}^\alpha_{l\sigma} c^\pm_{l-\sigma} c^\mp_{i-\sigma} c_{i\sigma}; c^+_{j\sigma} \rangle\rangle_\varepsilon \; . \tag{24}$$

The function $\langle\langle \hat{n}^\alpha_{l\sigma} c^\pm_{l-\sigma} c^\mp_{i-\sigma} c_{i\sigma}; c^+_{j\sigma} \rangle\rangle_\varepsilon$ fulfills the following equation of motion

$$\begin{aligned}
\varepsilon \langle\langle \hat{n}^\alpha_{l\sigma} c^\pm_{l-\sigma} c^\mp_{i-\sigma} c_{i\sigma}; c^+_{j\sigma} \rangle\rangle_\varepsilon &= \delta_{ij} \langle \hat{n}^\alpha_{l\sigma} c^\pm_{l-\sigma} c^\mp_{i-\sigma} \rangle - \delta_{jl} \xi_\alpha \langle c^+_{l\sigma} c^\pm_{l-\sigma} c^\mp_{i-\sigma} c_{i\sigma} \rangle \\
&- \xi_\alpha \sum_m t_{ml} \left( \langle\langle c^+_{l\sigma} c_{m\sigma} c^\pm_{l-\sigma} c^\mp_{i-\sigma} c_{i\sigma}; c^+_{j\sigma} \rangle\rangle_\varepsilon - \langle\langle c^+_{m\sigma} c_{l\sigma} c^\pm_{l-\sigma} c^\mp_{i-\sigma} c_{i\sigma}; c^+_{j\sigma} \rangle\rangle_\varepsilon \right) \\
&\pm \sum_m t_{ml} \langle\langle \hat{n}^\alpha_{l\sigma} c^\pm_{m-\sigma} c^\mp_{i-\sigma} c_{i\sigma}; c^+_{j\sigma} \rangle\rangle_\varepsilon \mp \sum_m t_{im} \langle\langle \hat{n}^\alpha_{l\sigma} c^\pm_{l-\sigma} c^\mp_{m-\sigma} c_{i\sigma}; c^+_{j\sigma} \rangle\rangle_\varepsilon \\
&- \sum_m t_{im} \langle\langle \hat{n}^\alpha_{l\sigma} c^\pm_{l-\sigma} c^\mp_{i-\sigma} c_{m\sigma}; c^+_{j\sigma} \rangle\rangle_\varepsilon + (\varepsilon_\pm \pm \varepsilon_- \mp \varepsilon_\alpha) \langle\langle \hat{n}^\alpha_{l\sigma} c^\pm_{l-\sigma} c^\mp_{i-\sigma} c_{i\sigma}; c^+_{j\sigma} \rangle\rangle_\varepsilon
\end{aligned} \tag{25}$$

For the moment we will consider only functions with the upper indices above

$$\begin{aligned}
\varepsilon \langle\langle \hat{n}^\alpha_{l\sigma} c^+_{l-\sigma} c^-_{i-\sigma} c_{i\sigma}; c^+_{j\sigma} \rangle\rangle_\varepsilon &= \delta_{ij} \langle \hat{n}^\alpha_{l\sigma} c^+_{l-\sigma} c^-_{i-\sigma} \rangle - \delta_{jl} \xi_\alpha \langle c^+_{l\sigma} c^+_{l-\sigma} c^-_{i-\sigma} c_{i\sigma} \rangle \\
&- \xi_\alpha \sum_m t_{ml} \left( \langle\langle c^+_{l\sigma} c_{m\sigma} c^+_{l-\sigma} c^-_{i-\sigma} c_{i\sigma}; c^+_{j\sigma} \rangle\rangle_\varepsilon - \langle\langle c^+_{m\sigma} c_{l\sigma} c^+_{l-\sigma} c^-_{i-\sigma} c_{i\sigma}; c^+_{j\sigma} \rangle\rangle_\varepsilon \right) \\
&+ \sum_m t_{ml} \langle\langle \hat{n}^\alpha_{l\sigma} c^+_{m-\sigma} c^-_{i-\sigma} c_{i\sigma}; c^+_{j\sigma} \rangle\rangle_\varepsilon - \sum_m t_{im} \langle\langle \hat{n}^\alpha_{l\sigma} c^+_{l-\sigma} c^-_{m-\sigma} c_{i\sigma}; c^+_{j\sigma} \rangle\rangle_\varepsilon \\
&- \sum_m t_{im} \langle\langle \hat{n}^\alpha_{l\sigma} c^+_{l-\sigma} c^-_{i-\sigma} c_{m\sigma}; c^+_{j\sigma} \rangle\rangle_\varepsilon + (\varepsilon_+ + \varepsilon_- - \varepsilon_\alpha) \langle\langle \hat{n}^\alpha_{l\sigma} c^+_{l-\sigma} c^-_{i-\sigma} c_{i\sigma}; c^+_{j\sigma} \rangle\rangle_\varepsilon
\end{aligned} \tag{26}$$

For the Green functions appearing in Eq. (26) we will use the following approximations

$$\begin{aligned}
\langle\langle c^+_{l\sigma} c_{m\sigma} c^+_{l-\sigma} c^-_{i-\sigma} c_{i\sigma}; c^+_{j\sigma} \rangle\rangle_\varepsilon &\cong \langle c^+_{l\sigma} c_{m\sigma} \rangle \langle\langle c^+_{l-\sigma} c^-_{i-\sigma} c_{i\sigma}; c^+_{j\sigma} \rangle\rangle_\varepsilon - \langle c^+_{l\sigma} c_{i\sigma} \rangle \langle\langle c^+_{l-\sigma} c^-_{i-\sigma} c_{m\sigma}; c^+_{j\sigma} \rangle\rangle_\varepsilon \\
&+ \langle c^+_{l\sigma} c_{m\sigma} c^+_{l-\sigma} c^-_{i-\sigma} \rangle \langle\langle c_{i\sigma}; c^+_{j\sigma} \rangle\rangle_\varepsilon - \langle c^+_{l\sigma} c_{i\sigma} c^+_{l-\sigma} c^-_{i-\sigma} \rangle \langle\langle c_{m\sigma}; c^+_{j\sigma} \rangle\rangle_\varepsilon \\
&\cong - \langle c^+_{l\sigma} c_{i\sigma} c^+_{l-\sigma} c^-_{i-\sigma} \rangle \langle\langle c_{m\sigma}; c^+_{j\sigma} \rangle\rangle_\varepsilon
\end{aligned} \tag{27}$$

$$\begin{aligned}
\langle\langle c^+_{m\sigma} c_{l\sigma} c^+_{l-\sigma} c^-_{i-\sigma} c_{i\sigma}; c^+_{j\sigma} \rangle\rangle_\varepsilon &\cong \langle c^+_{m\sigma} c_{l\sigma} \rangle \langle\langle c^+_{l-\sigma} c^-_{i-\sigma} c_{i\sigma}; c^+_{j\sigma} \rangle\rangle_\varepsilon - \langle c^+_{m\sigma} c_{i\sigma} \rangle \langle\langle c^+_{l-\sigma} c^-_{i-\sigma} c_{l\sigma}; c^+_{j\sigma} \rangle\rangle_\varepsilon \\
&+ \langle c^+_{m\sigma} c_{l\sigma} c^+_{l-\sigma} c^-_{i-\sigma} \rangle \langle\langle c_{i\sigma}; c^+_{j\sigma} \rangle\rangle_\varepsilon - \langle c^+_{m\sigma} c_{i\sigma} c^+_{l-\sigma} c^-_{i-\sigma} \rangle \langle\langle c_{l\sigma}; c^+_{j\sigma} \rangle\rangle_\varepsilon \cong 0
\end{aligned} \tag{28}$$

$$\begin{aligned}
\langle\langle \hat{n}^\alpha_{l\sigma} c^+_{m-\sigma} c^-_{i-\sigma} c_{i\sigma}; c^+_{j\sigma} \rangle\rangle_\varepsilon &\cong n^\alpha_\sigma \langle\langle c^+_{m-\sigma} c^-_{i-\sigma} c_{i\sigma}; c^+_{j\sigma} \rangle\rangle_\varepsilon - \xi_\alpha \langle c^+_{l\sigma} c_{i\sigma} \rangle \langle\langle c^+_{m-\sigma} c^-_{i-\sigma} c_{l\sigma}; c^+_{j\sigma} \rangle\rangle_\varepsilon \\
&+ \langle \hat{n}^\alpha_{l\sigma} c^+_{m-\sigma} c^-_{i-\sigma} \rangle \langle\langle c_{i\sigma}; c^+_{j\sigma} \rangle\rangle_\varepsilon - \xi_\alpha \langle c^+_{l\sigma} c_{i\sigma} c^+_{m-\sigma} c^-_{i-\sigma} \rangle \langle\langle c_{l\sigma}; c^+_{j\sigma} \rangle\rangle_\varepsilon \cong n^\alpha_\sigma \langle\langle c^+_{m-\sigma} c^-_{i-\sigma} c_{i\sigma}; c^+_{j\sigma} \rangle\rangle_\varepsilon
\end{aligned} \tag{29}$$

$$\langle\langle \hat{n}^{\alpha}_{l\sigma} c^{+}_{l-\sigma} c^{-}_{m-\sigma} c_{i\sigma}; c^{+}_{j\sigma} \rangle\rangle_{\varepsilon} \cong n^{\alpha}_{\sigma}\langle\langle c^{+}_{l-\sigma} c^{-}_{m-\sigma} c_{i\sigma}; c^{+}_{j\sigma} \rangle\rangle_{\varepsilon} - \xi_{\alpha}\langle c^{+}_{l\sigma} c_{i\sigma}\rangle\langle\langle c^{+}_{l-\sigma} c^{-}_{m-\sigma} c_{l\sigma}; c^{+}_{j\sigma} \rangle\rangle_{\varepsilon}$$
$$+\langle \hat{n}^{\alpha}_{l\sigma} c^{+}_{l-\sigma} c^{-}_{m-\sigma}\rangle\langle\langle c_{i\sigma}; c^{+}_{j\sigma}\rangle\rangle_{\varepsilon} - \xi_{\alpha}\langle c^{+}_{l\sigma} c_{i\sigma} c^{+}_{l-\sigma} c^{-}_{m-\sigma}\rangle\langle\langle c_{l\sigma}; c^{+}_{j\sigma}\rangle\rangle_{\varepsilon} \cong \delta_{lm} n^{\alpha}_{\sigma} n^{+}_{-\sigma} \langle\langle c_{i\sigma}; c^{+}_{j\sigma}\rangle\rangle_{\varepsilon}$$ (30)

$$\langle\langle \hat{n}^{\alpha}_{l\sigma} c^{+}_{l-\sigma} c^{-}_{i-\sigma} c_{m\sigma}; c^{+}_{j\sigma} \rangle\rangle_{\varepsilon} \cong n^{\alpha}_{\sigma}\langle\langle c^{+}_{l-\sigma} c^{-}_{i-\sigma} c_{m\sigma}; c^{+}_{j\sigma}\rangle\rangle_{\varepsilon} - \xi_{\alpha}\langle c^{+}_{l\sigma} c_{m\sigma}\rangle\langle\langle c^{+}_{l-\sigma} c^{-}_{i-\sigma} c_{l\sigma}; c^{+}_{j\sigma}\rangle\rangle_{\varepsilon}$$
$$+\langle \hat{n}^{\alpha}_{l\sigma} c^{+}_{l-\sigma} c^{-}_{i-\sigma}\rangle\langle\langle c_{m\sigma}; c^{+}_{j\sigma}\rangle\rangle_{\varepsilon} - \xi_{\alpha}\langle c^{+}_{l\sigma} c_{m\sigma} c^{+}_{l-\sigma} c^{-}_{i-\sigma}\rangle\langle\langle c_{l\sigma}; c^{+}_{j\sigma}\rangle\rangle_{\varepsilon} \cong \langle \hat{n}^{\alpha}_{l\sigma} c^{+}_{l-\sigma} c^{-}_{i-\sigma}\rangle\langle\langle c_{m\sigma}; c^{+}_{j\sigma}\rangle\rangle_{\varepsilon}$$ (31)

We apply the same procedure to the case of lower indices in Eq. (25). After inserting both these sets of equations into Eq. (25) we arrive at the relation

$$[\varepsilon - (\varepsilon_{\pm} \pm \varepsilon_{-} \mp \varepsilon_{\alpha})]\langle\langle \hat{n}^{\alpha}_{l\sigma} c^{\pm}_{l-\sigma} c^{\mp}_{i-\sigma} c_{i\sigma}; c^{+}_{j\sigma}\rangle\rangle_{\varepsilon} = \langle \hat{n}^{\alpha}_{l\sigma} c^{\pm}_{l-\sigma} c^{\mp}_{i-\sigma}\rangle\left(\delta_{ij} - \sum_{m} t_{im}\langle\langle c_{m\sigma}; c^{+}_{j\sigma}\rangle\rangle_{\varepsilon}\right)$$
$$-\xi_{\alpha}\langle c^{+}_{l\sigma} c_{i\sigma} c^{\pm}_{l-\sigma} c^{\mp}_{i-\sigma}\rangle\left(\delta_{jl} - \sum_{m} t_{lm}\langle\langle c_{m\sigma}; c^{+}_{j\sigma}\rangle\rangle_{\varepsilon}\right) \pm n^{\alpha}_{\sigma}\sum_{m} t_{ml}\langle\langle c^{\pm}_{m-\sigma} c^{\mp}_{i-\sigma} c_{i\sigma}; c^{+}_{j\sigma}\rangle\rangle_{\varepsilon}$$ (32)
$$\mp t_{il} n^{\alpha}_{\sigma} n^{\pm}_{-\sigma}\langle\langle c_{i\sigma}; c^{+}_{j\sigma}\rangle\rangle_{\varepsilon}$$

Dividing both sides by $\varepsilon - (\varepsilon_{\pm} \pm \varepsilon_{-} \mp \varepsilon_{\alpha})$ and using Eq. (16) and one can write that

$$\langle\langle \hat{n}^{\alpha}_{l\sigma} c^{\pm}_{l-\sigma} c^{\mp}_{i-\sigma} c_{i\sigma}; c^{+}_{j\sigma}\rangle\rangle_{\varepsilon} = \frac{\langle \hat{n}^{\alpha}_{l\sigma} c^{\pm}_{l-\sigma} c^{\mp}_{i-\sigma}\rangle}{\varepsilon - (\varepsilon_{\pm} \pm \varepsilon_{-} \mp \varepsilon_{\alpha})} F^{\sigma}_{H,0}(\varepsilon)\langle\langle c_{i\sigma}; c^{+}_{j\sigma}\rangle\rangle_{\varepsilon}$$
$$\pm \frac{n^{\alpha}_{\sigma}}{\varepsilon - (\varepsilon_{\pm} \pm \varepsilon_{-} \mp \varepsilon_{\alpha})}\left[\sum_{m} t_{ml}\langle\langle c^{\pm}_{m-\sigma} c^{\mp}_{i-\sigma} c_{i\sigma}; c^{+}_{j\sigma}\rangle\rangle_{\varepsilon} - t_{il} n^{\pm}_{-\sigma}\langle\langle c_{i\sigma}; c^{+}_{j\sigma}\rangle\rangle_{\varepsilon}\right].$$ (33)
$$-\frac{\xi_{\alpha}}{\varepsilon - (\varepsilon_{\pm} \pm \varepsilon_{-} \mp \varepsilon_{\alpha})}\langle c^{+}_{l\sigma} c_{i\sigma} c^{\pm}_{l-\sigma} c^{\mp}_{i-\sigma}\rangle F^{\sigma}_{H,0}(\varepsilon)\langle\langle c_{l\sigma}; c^{+}_{j\sigma}\rangle\rangle_{\varepsilon}$$

Summing up over $\alpha = \pm$ we arrive at the relation

$$\langle\langle c^{\pm}_{l-\sigma} c^{\mp}_{i-\sigma} c_{i\sigma}; c^{+}_{j\sigma}\rangle\rangle_{\varepsilon} = \sum_{\alpha} \frac{\langle \hat{n}^{\alpha}_{l\sigma} c^{\pm}_{l-\sigma} c^{\mp}_{i-\sigma}\rangle}{\varepsilon - (\varepsilon_{\pm} \pm \varepsilon_{-} \mp \varepsilon_{\alpha})} F^{\sigma}_{H,0}(\varepsilon)\langle\langle c_{i\sigma}; c^{+}_{j\sigma}\rangle\rangle_{\varepsilon}$$
$$-\frac{1}{F^{-\sigma}_{H,0}(\varepsilon_{-} \pm \varepsilon_{\pm} \mp \varepsilon)}\left[\sum_{m} t_{ml}\langle\langle c^{\pm}_{m-\sigma} c^{\mp}_{i-\sigma} c_{i\sigma}; c^{+}_{j\sigma}\rangle\rangle_{\varepsilon} - t_{il} n^{\pm}_{-\sigma}\langle\langle c_{i\sigma}; c^{+}_{j\sigma}\rangle\rangle_{\varepsilon}\right].$$ (34)
$$-\sum_{\alpha} \frac{\xi_{\alpha}}{\varepsilon - (\varepsilon_{\pm} \pm \varepsilon_{-} \mp \varepsilon_{\alpha})}\langle c^{+}_{l\sigma} c_{i\sigma} c^{\pm}_{l-\sigma} c^{\mp}_{i-\sigma}\rangle F^{\sigma}_{H,0}(\varepsilon)\langle\langle c_{l\sigma}; c^{+}_{j\sigma}\rangle\rangle_{\varepsilon}$$

The term in the square brackets above is the Hubbard solution.[3] Two extra terms are the corrections for the inter-site correlations. Solution to Eq. (34) will be the sum of Hubbard solution and the inter-site correction, and it will take on the following form

$$\langle\langle c^{\pm}_{l-\sigma} c^{\mp}_{i-\sigma} c_{i\sigma}; c^{+}_{j\sigma}\rangle\rangle_{\varepsilon} = -\sum_{m} W^{-\sigma}_{lm,i}(\varepsilon_{-} \pm \varepsilon_{\pm} \mp \varepsilon) t_{mi}\langle\langle (n^{\pm}_{i-\sigma} - n^{\pm}_{-\sigma}) c_{i\sigma}; c^{+}_{j\sigma}\rangle\rangle_{\varepsilon}$$
$$+\sum_{\alpha} \frac{\langle \hat{n}^{\alpha}_{l\sigma} c^{\pm}_{l-\sigma} c^{\mp}_{i-\sigma}\rangle}{\varepsilon - (\varepsilon_{\pm} \pm \varepsilon_{-} \mp \varepsilon_{\alpha})} F^{\sigma}_{H,0}(\varepsilon)\langle\langle c_{i\sigma}; c^{+}_{j\sigma}\rangle\rangle_{\varepsilon} - \sum_{\alpha} \frac{\xi_{\alpha}\langle c^{+}_{l\sigma} c_{i\sigma} c^{\pm}_{l-\sigma} c^{\mp}_{i-\sigma}\rangle}{\varepsilon - (\varepsilon_{\pm} \pm \varepsilon_{-} \mp \varepsilon_{\alpha})} F^{\sigma}_{H,0}(\varepsilon)\langle\langle c_{l\sigma}; c^{+}_{j\sigma}\rangle\rangle_{\varepsilon}$$ , (35)

where the last two extra terms are responsible for the inter-site averages.

Inserting to Eq. (7) the Green function from Eq. (35), still ignoring in Eq. (7) the terms with Green function $\langle\langle(\hat{n}^{\alpha}_{i-\sigma}-n^{\alpha}_{-\sigma})c_{l\sigma};c^+_{j\sigma}\rangle\rangle_{\varepsilon}$ describing the scattering correction, one obtains for the Green functions: $\langle\langle\hat{n}^+_{i-\sigma}c_{i\sigma};c^+_{j\sigma}\rangle\rangle_{\varepsilon}$, and $\langle\langle\hat{n}^-_{i-\sigma}c_{i\sigma};c^+_{j\sigma}\rangle\rangle_{\varepsilon}$, the matrix equation

$$\begin{bmatrix} \varepsilon-\varepsilon_- - n^+_{-\sigma}\Omega^B_{-\sigma}(\varepsilon) & n^-_{-\sigma}\Omega^B_{-\sigma}(\varepsilon) \\ n^+_{-\sigma}\Omega^B_{-\sigma}(\varepsilon) & \varepsilon-\varepsilon_+ - n^-_{-\sigma}\Omega^B_{-\sigma}(\varepsilon) \end{bmatrix}\begin{bmatrix} \langle\langle\hat{n}^-_{i-\sigma}c_{i\sigma};c^+_{j\sigma}\rangle\rangle_{\varepsilon} \\ \langle\langle\hat{n}^+_{i-\sigma}c_{i\sigma};c^+_{j\sigma}\rangle\rangle_{\varepsilon} \end{bmatrix}$$
$$= \begin{bmatrix} \hat{n}^-_{i-\sigma} \\ \hat{n}^+_{i-\sigma} \end{bmatrix}\left[\delta_{ij} - \sum_l t_{il}\langle\langle c_{l\sigma};c^+_{j\sigma}\rangle\rangle_{\varepsilon}\right] + \begin{bmatrix} -1 \\ +1 \end{bmatrix}\left\{\sum_l B^B_{il,\sigma}(\varepsilon)\langle\langle c_{l\sigma};c^+_{j\sigma}\rangle\rangle_{\varepsilon} + S^B_{\sigma}(\varepsilon)\langle\langle c_{i\sigma};c^+_{j\sigma}\rangle\rangle_{\varepsilon}\right\}, \quad (36)$$

where

$$\Omega^B_{-\sigma}(\varepsilon) = \sum_{lm} t_{il}t_{mi}\left[W^{\sigma}_{lm,i}(\varepsilon) - W^{\sigma}_{lm,i}(\varepsilon_-+\varepsilon_+-\varepsilon)\right] = \Omega_{-\sigma}(\varepsilon) - \Omega_{-\sigma}(\varepsilon_-+\varepsilon_+-\varepsilon), \quad (37)$$

$$B^B_{il,\sigma}(\varepsilon) = -t_{il}F^{\sigma}_{H,0}(\varepsilon)C_-(\varepsilon)\langle c^+_{l\sigma}c_{i\sigma}(c_{l-\sigma}c^+_{i-\sigma}-c^+_{l-\sigma}c_{i-\sigma})\rangle, \quad (38)$$

$$S^B_{\sigma}(\varepsilon) = F^{\sigma}_{H,0}(\varepsilon)C_-(\varepsilon)\sum_{l,i}(-t_{il})\left(2\langle\hat{n}_{l\sigma}c^+_{l-\sigma}c_{i-\sigma}\rangle - \langle c^+_{l-\sigma}c_{i-\sigma}\rangle\right). \quad (39)$$

**Scattering and resonance broadening effects together**

Finally, the scattering effect will be now combined with the resonance broadening effect. Equation (21) for scattering effect and equation (36) for resonance broadening effect have the same form. Therefore they can be written as one equation in the matrix form

$$\begin{bmatrix} \varepsilon-\varepsilon_- - n^+_{-\sigma}\Omega^T_{-\sigma}(\varepsilon) & n^-_{-\sigma}\Omega^T_{-\sigma}(\varepsilon) \\ n^+_{-\sigma}\Omega^T_{-\sigma}(\varepsilon) & \varepsilon-\varepsilon_+ - n^-_{-\sigma}\Omega^T_{-\sigma}(\varepsilon) \end{bmatrix}\begin{bmatrix} \langle\langle\hat{n}^-_{i-\sigma}c_{i\sigma};c^+_{j\sigma}\rangle\rangle_{\varepsilon} \\ \langle\langle\hat{n}^+_{i-\sigma}c_{i\sigma};c^+_{j\sigma}\rangle\rangle_{\varepsilon} \end{bmatrix}$$
$$= \begin{bmatrix} \hat{n}^-_{i-\sigma} \\ \hat{n}^+_{i-\sigma} \end{bmatrix}\left[\delta_{ij} - \sum_l t_{il}\langle\langle c_{l\sigma};c^+_{j\sigma}\rangle\rangle_{\varepsilon}\right] + \begin{bmatrix} -1 \\ +1 \end{bmatrix}\sum_l\left\{B^S_{il,\sigma}(\varepsilon) + B^B_{il,\sigma}(\varepsilon)\right\}\langle\langle c_{l\sigma};c^+_{j\sigma}\rangle\rangle_{\varepsilon}, \quad (40)$$
$$+ \begin{bmatrix} -1 \\ +1 \end{bmatrix}S^B_{\sigma}(\varepsilon)\langle\langle c_{i\sigma};c^+_{j\sigma}\rangle\rangle_{\varepsilon}$$

where

$$\Omega^T_{\sigma}(\varepsilon) = \Omega_{\sigma}(\varepsilon) + \Omega_{-\sigma}(\varepsilon) - \Omega_{-\sigma}(\varepsilon_-+\varepsilon_+-\varepsilon). \quad (41)$$

Solving Eq. (40) we find the functions $\langle\langle\hat{n}^+_{i-\sigma}c_{i\sigma};c^+_{j\sigma}\rangle\rangle_{\varepsilon}$ and $\langle\langle\hat{n}^-_{i-\sigma}c_{i\sigma};c^+_{j\sigma}\rangle\rangle_{\varepsilon}$, and after using the identity

$$\sum_{\alpha=\pm}\langle\langle\hat{n}^{\alpha}_{i-\sigma}c_{i\sigma};c^+_{j\sigma}\rangle\rangle_{\varepsilon} = \langle\langle c_{i\sigma};c^+_{j\sigma}\rangle\rangle_{\varepsilon} = G^{\sigma}_{ij}(\varepsilon), \quad (42)$$

we have

$$F^{\sigma}_H(\varepsilon)G^{\sigma}_{ij}(\varepsilon) = \delta_{ij} - \sum_l t_{il}G^{\sigma}_{lj}(\varepsilon) + \sum_l B^T_{il,\sigma}(\varepsilon)G^{\sigma}_{lj}(\varepsilon) + S^T_{\sigma}(\varepsilon)G^{\sigma}_{ij}(\varepsilon), \quad (43)$$

where

$$B_{il,\sigma}^T(\varepsilon) = F_H^\sigma(\varepsilon)C_-[\varepsilon - \Omega_\sigma^T(\varepsilon)]F_{H,0}^\sigma(\varepsilon)C_-(\varepsilon)(-t_{il})$$
$$[\langle \hat{n}_{i-\sigma}\hat{n}_{l-\sigma}\rangle - n_{-\sigma}^2 + \langle c_{l\sigma}^+ c_{i\sigma}(c_{l-\sigma}c_{i-\sigma}^+ - c_{l-\sigma}^+ c_{i-\sigma})\rangle] \quad , \tag{44}$$

$$S_\sigma^T(\varepsilon) = F_H^\sigma(\varepsilon)C_-[\varepsilon - \Omega_\sigma^T(\varepsilon)]S_\sigma^B(\varepsilon)$$
$$= F_H^\sigma(\varepsilon)C_-[\varepsilon - \Omega_\sigma^T(\varepsilon)]F_{H,0}^\sigma(\varepsilon)C_-(\varepsilon)\sum_{li}(-t_{il})(2\langle \hat{n}_{l\sigma}c_{-\sigma}^+ c_{i-\sigma}\rangle - \langle c_{l-\sigma}^+ c_{i-\sigma}\rangle) \quad , \tag{45}$$

$$\frac{1}{F_H^\sigma(\varepsilon)} = \frac{\varepsilon - (n_{-\sigma}^+\varepsilon_- + n_{-\sigma}^-\varepsilon_+) - \Omega_\sigma^T(\varepsilon)}{[\varepsilon - \varepsilon_- - n_{-\sigma}^+\Omega_\sigma^T(\varepsilon)][\varepsilon - \varepsilon_+ - n_{-\sigma}^-\Omega_\sigma^T(\varepsilon)] - n_{-\sigma}^- n_{-\sigma}^+[\Omega_\sigma^T(\varepsilon)]^2} \quad . \tag{46}$$

Equation (43) is solved by applying Fourier transformation to the momentum space and using the relation

$$\sum_l B_{il,\sigma}^T(\varepsilon)G_{lj}^\sigma(\varepsilon) = \sum_k B_{k,\sigma}^T(\varepsilon)G_k^\sigma(\varepsilon)\exp[i\mathbf{k}\cdot(\mathbf{r}_i - \mathbf{r}_j)] \quad , \tag{47}$$

where

$$B_{k,\sigma}^T(\varepsilon) = F_H^\sigma(\varepsilon)C_-[\varepsilon - \Omega_\sigma^T(\varepsilon)]F_{H,0}^\sigma(\varepsilon)C_-(\varepsilon)\frac{1}{N}\sum_{il}(-t_{il})(\langle \hat{n}_{l-\sigma}\hat{n}_{i-\sigma}\rangle - n_{-\sigma}^2$$
$$-\langle c_{l\sigma}^+ c_{i-\sigma}^+ c_{l-\sigma}c_{i\sigma}\rangle - \langle c_{l\sigma}^+ c_{l-\sigma}^+ c_{i-\sigma}c_{i\sigma}\rangle)\exp[i\mathbf{k}\cdot(\mathbf{r}_l - \mathbf{r}_i)] \quad . \tag{48}$$

As a result we obtain from Eq. (43) the following form

$$F_H^\sigma(\varepsilon)\sum_k G_k^\sigma(\varepsilon)\exp[i\mathbf{k}\cdot(\mathbf{r}_i - \mathbf{r}_j)] = \sum_k [1 + \varepsilon_k G_k^\sigma(\varepsilon)]\exp[i\mathbf{k}\cdot(\mathbf{r}_i - \mathbf{r}_j)]$$
$$+ \sum_k [B_{k,\sigma}^T(\varepsilon) + S_\sigma^T(\varepsilon)]G_k^\sigma(\varepsilon)\exp[i\mathbf{k}\cdot(\mathbf{r}_i - \mathbf{r}_j)] \quad , \tag{49}$$

from which we arrive at the final result

$$G_k^\sigma(\varepsilon) = \frac{1}{F_H^\sigma(\varepsilon) - \varepsilon_k - [B_{k,\sigma}^T(\varepsilon) + S_\sigma^T(\varepsilon)]} \quad . \tag{50}$$

This Green function is the final solution of Eq. (7), which contains the scattering and resonance broadening effects which include the inter-site correlations.

The bandshift correction factor $S_\sigma^T(\varepsilon)$ causes exchange splitting between the spin-up and spin-down spectrum, and the bandwidth correction factor $B_{k,\sigma}^T(\varepsilon)$ leads to a change in the width of the spin sub-bands with respect to each other.

Assuming that $B_{k,\sigma}^T(\varepsilon) = 0$ and $S_\sigma^T(\varepsilon) = 0$ in the Eq. (50) we obtain the classic Hubbard III approximation

$$G_k^\sigma(\varepsilon) = \frac{1}{F_H^\sigma(\varepsilon) - \varepsilon_k} \quad . \tag{51}$$

As is now well known, the standard Hubbard III approximation (both scattering and resonance broadening effects) is equivalent to the CPA approximation under the following change of variables between the Hubbard solution and the CPA approximation[6]

$$F_H^\sigma(\varepsilon) \to \varepsilon - \Sigma^\sigma(\varepsilon)$$
$$n_{-\sigma}^- \to P_1^\sigma, \quad \varepsilon_- \to V_1$$
$$n_{-\sigma}^+ \to P_2^\sigma, \quad \varepsilon_+ \to V_2 \quad , \qquad (52)$$
$$U \equiv \varepsilon_+ - \varepsilon_- \to V_2 - V_1$$
$$\Omega_\sigma^T(\varepsilon) \to \varepsilon - \Sigma^\sigma(\varepsilon) - \frac{1}{F^\sigma(\varepsilon)}$$

where $F^\sigma(\varepsilon)$ is the Slater-Koster function defined as

$$F^\sigma(\varepsilon) = \frac{1}{N} \sum_k G_k^\sigma(\varepsilon) \quad . \qquad (53)$$

Under this change of variables, it is straightforward to demonstrate that the Hubbard's relations (51) and (46) above are identical to the CPA relations:

$$G_k^\sigma(\varepsilon) = \frac{1}{\varepsilon - \varepsilon_k - \Sigma^\sigma(\varepsilon)} \quad , \qquad (54)$$

and

$$\Sigma^\sigma(\varepsilon) = \bar{\varepsilon}^\sigma + \frac{(V_2 - V_1)^2 P_1^\sigma P_2^\sigma F^\sigma(\varepsilon)}{1 + [\Sigma^\sigma(\varepsilon) + \bar{\varepsilon}^\sigma - (V_1 + V_2)]F^\sigma(\varepsilon)} \quad , \qquad (55)$$

with potentials and probabilities given by

$$V_i = \begin{cases} 0 \\ U \end{cases}, \qquad P_i^\sigma = \begin{cases} 1 - n_{i-\sigma} \\ n_{i-\sigma} \end{cases}, \qquad (56)$$

and $\bar{\varepsilon}^\sigma = P_1^\sigma V_1 + P_2^\sigma V_2$ .

It has been shown (see Refs. 28 and 29) that the CPA approximation [Eq. (55)] with potentials and probabilities given by Eq. (56) does not bring about the ferromagnetic ground state. Therefore, the above identification of CPA with Hubbard III approximation proves that the Hubbard III approximation is also not ferromagnetic.

### IV. FORMALISM FOR CALCULATING THE DENSITY OF STATES

In this Section we will introduce general formulas for calculating the density of states (DOS), the Slater-Koster function and the average electron occupation number.

The DOS for electrons with spin $\sigma$ can be expressed as

$$\rho^{\sigma}(\varepsilon) = -\frac{1}{\pi} \operatorname{Im} F^{\sigma}(\varepsilon) \ , \tag{57}$$

where on the basis of Eqs. (50), (52) and (53) the Slater-Koster function $F^{\sigma}(\varepsilon)$ is given by

$$F^{\sigma}(\varepsilon) = \frac{1}{N} \sum_{k} \frac{1}{\varepsilon - \Sigma_{eff}^{\sigma}(\varepsilon) - \varepsilon_{k} b^{\sigma}(\varepsilon)} \ , \tag{58}$$

with

$$b^{\sigma}(\varepsilon) = 1 + \frac{B_{k,\sigma}^{T}(\varepsilon)}{\varepsilon_{k}}, \quad \text{and} \quad \Sigma_{eff}^{\sigma}(\varepsilon) = \Sigma^{\sigma}(\varepsilon) + S_{\sigma}^{T}(\varepsilon) \ . \tag{59}$$

According to Eqs. (58) and (59) the perturbed Slater-Koster function $F^{\sigma}(\varepsilon)$ and the perturbed DOS, will depend on the bandwidth factor $B_{k,\sigma}^{T}(\varepsilon)$ and the bandshift factor $S_{\sigma}^{T}(\varepsilon)$ given by Eqs. (48) and (45), respectively. Comparing Eq. (58) of the perturbed case with the unperturbed Slater-Koster function given by the relation

$$F_{0}(\varepsilon) = \frac{1}{N} \sum_{k} \frac{1}{\varepsilon - \varepsilon_{k}} = \int_{-\infty}^{\infty} \frac{\rho_{0}(\varepsilon')}{\varepsilon + i0^{+} - \varepsilon'} d\varepsilon' \ , \tag{60}$$

we can write

$$\rho^{\sigma}(\varepsilon) = -\frac{1}{\pi} \operatorname{Im} F^{\sigma}(\varepsilon) = -\frac{1}{\pi} \operatorname{Im} \left\{ \frac{1}{b^{\sigma}(\varepsilon)} F_{0}\left[\frac{\varepsilon - \Sigma_{eff}^{\sigma}(\varepsilon)}{b^{\sigma}(\varepsilon)}\right] \right\} \ . \tag{61}$$

The average electron occupation number can be found as

$$n_{\sigma} = \int_{-\infty}^{\infty} \rho^{\sigma}(\varepsilon) \frac{d\varepsilon}{e^{(\varepsilon-\mu)/k_{B}T} + 1} \ . \tag{62}$$

In the numerical calculations we will use the semi-elliptic unperturbed DOS given by

$$\rho_{0}(\varepsilon) = \frac{2}{\pi D^{2}} (D^{2} - \varepsilon^{2})^{1/2} \ , \tag{63}$$

where $D$ is the half bandwidth.

The shape of the DOS does not have a major impact on the result of calculations.

## V. THE BANDWIDTH AND BANDSHIFT CORRECTION

The bandwidth and bandshift corrections described by parameters $B_{k,\sigma}^{T}(\varepsilon)$ and $S_{\sigma}^{T}(\varepsilon)$, given by Eqs. (48) and (45), modify the Hubbard III solution.

Parameters $B^T_{k,\sigma}(\varepsilon)$ and $S^T_\sigma(\varepsilon)$ will be calculated either by applying to the correlation functions, appearing inside them, the Hartree-Fock approximation or the approximation developed by Roth[30] and Nolting and co-workers[32]

### A. Hartree-Fock solution for $B^T_{k,\sigma}(\varepsilon)$ and $S^T_\sigma(\varepsilon)$

In Eqs. (48) and (45) for $B^T_{k,\sigma}(\varepsilon)$ and $S^T_\sigma(\varepsilon)$ we assume for $l \neq i$ the following approximations

$$\langle \hat{n}_{l-\sigma} \hat{n}_{i-\sigma} \rangle - n^2_{-\sigma} \approx \langle c^+_{l-\sigma} c_{l-\sigma} \rangle \langle c^+_{i-\sigma} c_{i-\sigma} \rangle - \langle c^+_{l-\sigma} c_{i-\sigma} \rangle \langle c^+_{i-\sigma} c_{l-\sigma} \rangle - n^2_{-\sigma} = -I^2_{-\sigma}$$

$$\langle c^+_{l\sigma} c^+_{i-\sigma} c_{l-\sigma} c_{i\sigma} \rangle \approx \langle c^+_{l\sigma} c_{i\sigma} \rangle \langle c^+_{i-\sigma} c_{l-\sigma} \rangle = I_\sigma I_{-\sigma}$$

$$\langle c^+_{l\sigma} c^+_{l-\sigma} c_{i-\sigma} c_{i\sigma} \rangle \approx \langle c^+_{l\sigma} c_{i\sigma} \rangle \langle c^+_{l-\sigma} c_{i-\sigma} \rangle = I_\sigma I_{-\sigma} \quad (64)$$

$$\langle \hat{n}_{l\sigma} c^+_{l-\sigma} c_{i-\sigma} \rangle \approx \langle \hat{n}_{l\sigma} \rangle \langle c^+_{l-\sigma} c_{i-\sigma} \rangle = n_\sigma I_{-\sigma} \quad ,$$

where $I_\sigma = \langle c^+_{l\sigma} c_{i\sigma} \rangle$ is the Fock parameter.

We modified the standard Hartree-Fock approximation above by adding to the single-site two operator averages the inter-site two operator averages.

All these approximations reduce to zero when $I_\sigma = 0$. The approximation of the first expression in Eq. (64) to zero was put explicitly in the Hubbard paper.[3]
As a result of including Eq. (64) the expressions (48) and (45) for $B^T_{k,\sigma}(\varepsilon)$, and $S^T_\sigma(\varepsilon)$ will take on the forms

$$B^T_{k,\sigma}(\varepsilon) \approx -C_-[\varepsilon - \Omega^T_\sigma(\varepsilon)] F^\sigma_H(\varepsilon) C_-(\varepsilon) F^\sigma_{H,0}(\varepsilon) (I^2_{-\sigma} + 2 I_\sigma I_{-\sigma}) \varepsilon_k \quad , \quad (65)$$

$$S^T_\sigma(\varepsilon) \approx -C_-[\varepsilon - \Omega^T_\sigma(\varepsilon)] F^\sigma_H(\varepsilon) C_-(\varepsilon) F^\sigma_{H,0}(\varepsilon) D I_{-\sigma} (2 n_\sigma - 1) \quad . \quad (66)$$

According to its definition the parameter $I_\sigma$ is proportional to the average kinetic energy of electrons with spin $\sigma$

$$\langle K^\sigma \rangle = -t \left\langle \sum_{\langle il \rangle} c^+_{l\sigma} c_{i\sigma} \right\rangle = -tz \langle c^+_{l\sigma} c_{i\sigma} \rangle = -D I_\sigma \quad . \quad (67)$$

The average kinetic energy $\langle K^\sigma \rangle$ can be also written as

$$\langle K^\sigma \rangle = \int_{-D}^{D} \rho^\sigma(\varepsilon) \varepsilon f(\varepsilon) d\varepsilon \quad . \quad (68)$$

Comparing the above equations gives for the parameter $I_\sigma$ the relation

$$I_\sigma = \langle c^+_{l\sigma} c_{i\sigma} \rangle = \int_{-D}^{D} \rho^\sigma(\varepsilon)\left(-\frac{\varepsilon}{D}\right)\frac{d\varepsilon}{1+e^{(\varepsilon-\mu)/k_B T}}d\varepsilon \quad, \tag{69}$$

from which it can be directly calculated. It can also be approximated by its stochastic value, as the probability of electron hopping from the $j$-th to $i$-th lattice site. In the strong correlation case, $U \gg D$, the stochastic interpretation brings the following result for the lower Hubbard sub-band (see Gorski et. al.[37])

$$I_\sigma = \frac{n_\sigma(1-n)}{(1-n_{-\sigma})} \quad. \tag{70}$$

Using this relation and assuming $\varepsilon \approx 0$ for the lower Hubbard's sub-band in Eqs. (14) and (15) we can simplify the product $F^\sigma_{H,0}(\varepsilon)C_-(\varepsilon)$ in Eqs. (65) and (66) (in the case of $U \gg D$), and obtain

$$F^\sigma_{H,0}(\varepsilon)C_-(\varepsilon) = -\frac{1}{1-n_{-\sigma}} \quad. \tag{71}$$

Inserting this relation into Eqs. (65) and (66) one obtains

$$B^T_{k,\sigma}(\varepsilon) \equiv B^T_{k,\sigma} \approx -\frac{1}{(1-n_{-\sigma})^2}\left(I^2_{-\sigma} + 2I_\sigma I_{-\sigma}\right)\varepsilon_k \quad, \tag{72}$$

and

$$S^T_\sigma(\varepsilon) \equiv S^T_\sigma \approx \frac{D}{(1-n_{-\sigma})^2}I_{-\sigma}(1-2n_\sigma) \quad. \tag{73}$$

In the Hartree-Fock approximation the bandwidth factor $B^T_{k,\sigma}$ and the bandshift factor $S^T_\sigma$ are not dependent on the energy $\varepsilon$ therefore in this case the relation (62) can be simplified into the following form

$$n_\sigma = \int_{-\infty}^{\infty} \rho^\sigma(\varepsilon)\frac{d\varepsilon}{e^{(\varepsilon-\mu+S^T_\sigma)/k_B T}+1} \quad, \tag{74}$$

where the DOS is calculated from Eq. (61) with $b^\sigma$ real and independent on energy

$$\rho^\sigma(\varepsilon) = -\frac{1}{\pi}\operatorname{Im}F^\sigma(\varepsilon) = -\frac{1}{b^\sigma}\frac{1}{\pi}\operatorname{Im}F_0\left[\frac{\varepsilon-\Sigma^\sigma(\varepsilon)}{b^\sigma}\right] \quad, \tag{75}$$

and the bandwidth factor $b^\sigma$ related to $B^T_{k,\sigma}$ by Eq. (59) and given by

$$b^\sigma = 1 - \frac{1}{(1-n_{-\sigma})^2}\left(I^2_{-\sigma} + 2I_\sigma I_{-\sigma}\right) \quad. \tag{76}$$

Equations (70), (73)-(76) together with the CPA relation (55) and Eq. (60) build a closed system of equations, which has to be solved self-consistently.

**B. The high approximation for $B_{k,\sigma}^T(\varepsilon)$ and $S_\sigma^T(\varepsilon)$**

Now, the correlation functions appearing in the bandwidth correction factor $B_{k,\sigma}^T(\varepsilon)$ and in the bandshift correction factor $S_\sigma^T(\varepsilon)$ will be calculated based on the results of Nolting and Borgiel[32] and Roth.[30]

The Fock parameter $\langle c_{l-\sigma}^+ c_{i-\sigma} \rangle$ appearing in the bandshift correction factor $S_\sigma^T(\varepsilon)$ given by Eq. (45), can be written as

$$\langle c_{l-\sigma}^+ c_{i-\sigma} \rangle = \frac{1}{N} \sum_k \exp[i\mathbf{k} \cdot (\mathbf{r}_l - \mathbf{r}_i)] \int_{-\infty}^{\infty} S_{k-\sigma}(\varepsilon) \frac{d\varepsilon}{1+e^{(\varepsilon-\mu)/k_B T}} \quad (77)$$

where $S_{k-\sigma}(\varepsilon)$ is the single-electron spectral density

$$S_{k-\sigma}(\varepsilon) = -\frac{1}{\pi} \operatorname{Im} G_k^{-\sigma}(\varepsilon), \quad (78)$$

with the Green function given by Eq. (50).

The higher correlation function $\langle \hat{n}_{l\sigma} c_{l-\sigma}^+ c_{i-\sigma} \rangle$ [see Eq. (45)], can be expressed as (see Refs. 30 and 32)

$$\langle \hat{n}_{l\sigma} c_{l-\sigma}^+ c_{i-\sigma} \rangle = \frac{1}{UN} \sum_k \exp[i\mathbf{k} \cdot (\mathbf{r}_l - \mathbf{r}_i)] \int_{-\infty}^{\infty} (\varepsilon' - \varepsilon_k) S_{k-\sigma}(\varepsilon') \frac{d\varepsilon'}{1+e^{(\varepsilon'-\mu)/k_B T}}. \quad (79)$$

Inserting Eqs. (77) and (79) to the Eq. (45) we obtain finally

$$S_\sigma^T(\varepsilon) = F_H^\sigma(\varepsilon) C_-[\varepsilon - \Omega_\sigma^T(\varepsilon)] F_{H,0}^\sigma(\varepsilon) C_-(\varepsilon)$$
$$\times \frac{1}{N} \sum_k \int_{-\infty}^{\infty} \varepsilon_k \left[ \frac{2}{U} (\varepsilon' - \varepsilon_k) - 1 \right] S_{k-\sigma}(\varepsilon') \frac{d\varepsilon'}{1+e^{(\varepsilon'-\mu)/k_B T}}. \quad (80)$$

The three higher correlation functions: $\langle \hat{n}_{l-\sigma} \hat{n}_{i-\sigma} \rangle$, $\langle c_{l\sigma}^+ c_{l-\sigma}^+ c_{l-\sigma} c_{i\sigma} \rangle$, and $\langle c_{l\sigma}^+ c_{l-\sigma}^+ c_{i-\sigma} c_{i\sigma} \rangle$, which show up in the bandwidth correction factor $B_{k,\sigma}^T(\varepsilon)$ given by Eq. (48), can be written as[30,32]

$$\langle \hat{n}_{l-\sigma} \hat{n}_{i-\sigma} \rangle - n_{-\sigma}^2 = -\frac{\eta_{-\sigma} \langle c_{l-\sigma}^+ c_{i-\sigma} \rangle + \nu_{-\sigma} \langle \hat{n}_{l\sigma} c_{l-\sigma}^+ c_{i-\sigma} \rangle}{1+\nu_{0\sigma}\nu_{0-\sigma}}, \quad (81)$$

$$\langle c_{l\sigma}^+ c_{l-\sigma}^+ c_{l-\sigma} c_{i\sigma} \rangle = \frac{(\eta_\sigma + \nu_\sigma)\langle c_{l-\sigma}^+ c_{i-\sigma} \rangle - \nu_\sigma \langle \hat{n}_{l\sigma} c_{l-\sigma}^+ c_{i-\sigma} \rangle}{1-\nu_{0\sigma}}, \quad (82)$$

$$\langle c_{l\sigma}^+ c_{l-\sigma}^+ c_{i-\sigma} c_{i\sigma} \rangle = \frac{\eta_\sigma \langle c_{l-\sigma}^+ c_{i-\sigma} \rangle + \nu_\sigma \langle \hat{n}_{l\sigma} c_{l-\sigma}^+ c_{i-\sigma} \rangle}{1+\nu_{0\sigma}}, \quad (83)$$

where the following abbreviations were introduced:

$$\eta_\sigma = \frac{1}{1-n_{-\sigma}} \left( \langle c^+_{l-\sigma} c_{i-\sigma} \rangle - \langle \hat{n}_{l-\sigma} c^+_{l\sigma} c_{i\sigma} \rangle \right), \tag{84}$$

$$v_\sigma = \frac{1}{n_{-\sigma}(1-n_{-\sigma})} \left( \langle \hat{n}_{l-\sigma} c^+_{l\sigma} c_{i\sigma} \rangle - n_{-\sigma} \langle c^+_{l\sigma} c_{i\sigma} \rangle \right), \tag{85}$$

$$v_{0\sigma} = \frac{1}{n_{-\sigma}(1-n_{-\sigma})} \left( \langle \hat{n}_{l\sigma} \hat{n}_{l-\sigma} \rangle - n_\sigma n_{-\sigma} \right). \tag{86}$$

Inserting Eqs. (81)-(86) together with expressions (77) and (79) to Eq. (48) we obtain the bandwidth correction factor $B^T_{k,\sigma}(\varepsilon)$ as

$$B^T_{k,\sigma}(\varepsilon) = F^\sigma_H(\varepsilon) C_-[\varepsilon - \Omega^T_\sigma(\varepsilon)] F^\sigma_{H,0}(\varepsilon) C_-(\varepsilon) \frac{1}{N} \sum_{li} (-t_{li})$$
$$\left[ -\left( \frac{\eta_{-\sigma}}{1+v_{0\sigma}v_{0-\sigma}} + \frac{\eta_\sigma + v_\sigma}{1-v_{0\sigma}} + \frac{\eta_\sigma}{1+v_{0\sigma}} \right) \langle c^+_{l-\sigma} c_{i-\sigma} \rangle \right. \tag{87}$$
$$\left. -\left( \frac{v_{-\sigma}}{1+v_{0\sigma}v_{0-\sigma}} - \frac{v_\sigma}{1-v_{0\sigma}} + \frac{v_\sigma}{1+v_{0\sigma}} \right) \langle \hat{n}_{l\sigma} c^+_{l-\sigma} c_{i-\sigma} \rangle \right] \exp[i\mathbf{k} \cdot (\mathbf{r}_l - \mathbf{r}_i)]$$

The functions $F^\sigma_{H,0}(\varepsilon) C_-(\varepsilon)$ and $F^\sigma_H(\varepsilon) C_-[\varepsilon - \Omega^T_\sigma(\varepsilon)]$, appearing in expressions for $S^T_\sigma(\varepsilon)$ and $B^T_{k,\sigma}(\varepsilon)$, will with the help of Eqs. (14), (15) and (46) take on the following forms

$$F^\sigma_{H,0}(\varepsilon) C_-(\varepsilon) = \frac{U}{\varepsilon - U(1-n_{-\sigma})}, \tag{88}$$

and

$$F^\sigma_H(\varepsilon) C_-[\varepsilon - \Omega^T_\sigma(\varepsilon)] = \frac{U}{\varepsilon - \Omega^T_\sigma(\varepsilon) - U(1-n_{-\sigma})}. \tag{89}$$

The above two expressions have to be inserted back to formulas (80) for $S^T_\sigma(\varepsilon)$, and (87) for $B^T_{k,\sigma}(\varepsilon)$. Next, we use the expressions (80), (87) in Eq. (59) and the CPA Eq. (55). After calculating the self-energy $\Sigma^\sigma(\varepsilon)$ we use Eqs. (61) and (62) to calculate DOS and the average electron numbers with spin $\pm\sigma$.

### VI. MAGNETIC ANALYSIS OF THE SYSTEM GROUND STATE

In this Section the consequences of including: $B^T_{k,\sigma}(\varepsilon)$ and $S^T_\sigma(\varepsilon)$ for the appearance of the ferromagnetic ordering will be analyzed. The results will be compared here with the results of the standard CPA approach. In the next Section the comparison will be made with the similar results obtained for the strongly correlated systems by Nolting and co-workers, who also arrived at the magnetic ordered state using the SDA[31-33] and MAA theory.[34,35]

To analyze the possibility of the transition to ferromagnetic ground state we will add to the Hamiltonian (1) the term with the on-site atomic Stoner field $F_{in}$ in the H-F approximation

$$-\sum_{i\sigma} F_{in} n_\sigma \hat{n}_{i\sigma} \ . \tag{90}$$

After this extension the chemical potential in Eq. (62) will have the following form

$$\mu_\sigma = \mu + F_{in} n_\sigma \ . \tag{91}$$

In further analysis we will use two coupled equations for electron number and magnetization

$$n = n_\sigma + n_{-\sigma} \ , \qquad m = n_\sigma - n_{-\sigma} \ , \tag{92}$$

where $n_{\pm\sigma}$ is given by Eq. (62).

On the basis of these equations the critical on-site exchange interaction will be calculated in the limit of $m \to 0$. The cases of strong correlation, $U \gg D$, and intermediate correlation will be analyzed. The semi-elliptic initial band of Eq. (63) will be used.

The results will illustrate the role of the bandwidth factor $B_{k,\sigma}^T(\varepsilon)$ and the bandshift factor $S_\sigma^T(\varepsilon)$ in creating the ferromagnetic ground state.

In Figs. 1-3 we show the critical on-site exchange interaction $F_{in}^{cr}$ in the function of the electron occupation $n$ in the case of strong correlation, $U \gg D$. Fig. 1 presents the dependence $F_{in}^{cr}(n)$ with both the included factors: $B_{k,\sigma}^T(\varepsilon)$ and $S_\sigma^T(\varepsilon)$. Fig. 2 shows the influence of the bandshift factor $S_\sigma^T(\varepsilon)$ alone on $F_{in}^{cr}$, the bandwidth factor $B_{k,\sigma}^T(\varepsilon)=0$. Fig. 3 shows the influence of the bandwidth factor $B_{k,\sigma}^T(\varepsilon)$ alone on $F_{in}^{cr}$, the bandshift factor $S_\sigma^T(\varepsilon)=0$. It can be seen that the bandshift factor $S_\sigma^T(\varepsilon)$ (see Fig. 2) favors ferromagnetism much stronger than the bandwidth factor $B_{k,\sigma}^T(\varepsilon)$ (see Fig. 3). It has to be emphasized that in our model the bandshift factor comes from the inter-site correlation included within the Hubbard III approximation, not from the simple H-F approximation.

The curves $F_{in}^{cr}(n)$ in Figs. 1-3 are shown by three types of lines for all cases mentioned above. A solid line is the result based on a high approximation for factors $B_{k,\sigma}^T(\varepsilon)$ and $S_\sigma^T(\varepsilon)$ obtained in Sec. V.B, Eqs. (87) and (80). The dashed line is calculated using the H-F approximation for factors $B_{k,\sigma}^T(\varepsilon)$ and $S_\sigma^T(\varepsilon)$ derived in Sec. V.A, Eqs. (72) and (73). The dotted line is the classic CPA solution without the correlation factors. The self energy is calculated from Eq. (55).

Comparing the dashed and solid line with the CPA result in Fig. 1 one can see that the bandwidth factor $B_{k,\sigma}^T(\varepsilon)$ and the bandshift factor $S_\sigma^T(\varepsilon)$ significantly decreased the on-site exchange interaction necessary for ferromagnetism. In some concentration ranges, for which the critical on-site exchange field $F_{in}^{cr}$ drops below zero, these factors [$B_{k,\sigma}^T(\varepsilon)$ and $S_\sigma^T(\varepsilon)$] lead to ferromagnetic alignment coming from the interaction $U$, without adding the extra on-site exchange field. Maximum ferromagnetic enhancement takes place roughly in the middle of the lower Hubbard sub-band, for electron concentrations around $0.32 \leq n \leq 0.87$. All these figures show that the inter-site correlation factors effect ferromagnetism in the H-F approximation in a similar way to the high approximation.

We can see that the bandshift factor $S_\sigma^T(\varepsilon)$ is a dominant force behind the ferromagnetism. The influence of the bandwidth factor $B_{k,\sigma}^T(\varepsilon)$ is too weak to create ferromagnetism and the result is only the correction to the classic CPA in favor of ferromagnetism. The net result of both bandshift and bandwidth factors in the H-F and in the high approximation are close.

In Fig. 4 we show the critical on-site Stoner exchange field for different values of the Coulomb interaction $U$. Both the bandwidth and bandshift factors are included in the high approximation. One can see that the range of ferromagnetism is shrinking with decreasing $U$. It is also shifting towards higher concentrations. At the half-filled point all these curves match the corresponding CPA results,[38] as both correlation factors tend to zero in the limit of full sub-band. It has to be remembered that all values of $U$ used in our calculations cause split of the band, since the relation $U > D$ is fulfilled, and the lower sub-band becomes filed at $n = 1$.

In some concentration ranges the critical on-site Stoner exchange field is negative, meaning that we can expect spontaneous magnetization without this field. For these concentrations we calculated the Curie temperature $T_C$. The results are shown in Fig. 5. One can see a decreasing $T_C$ and the decreasing range of ferromagnetism with decreasing $U$.

## VII. DISCUSSION AND CONCLUSIONS

Including the inter-site kinetic correlation into the basic calculations of the Hubbard model[3] gives the solution, which at some values of electron occupation and Coulomb repulsion $U$ brings the ferromagnetic ground state. However, calculation of inter-site correlations appearing in Eqs. (45) and (48) is not fully self consistent and it is possible that introducing the requirement of self-consistency in this evaluation will eliminate the ferromagnetism, similarly as in the case of the CPA analysis.

Our results for $B_{k,\sigma}^T(\varepsilon)$ and $S_\sigma^T(\varepsilon)$, should be compared with the similar results obtained for the strongly correlated systems by Nolting and co-workers,[31-35] who have also arrived at the magnetic ordered state using the SDA and MAA theory. In the SDA method[31] the higher correlation function has been defined, which can be split as follows into a **k**-dependent and a **k**-independent term

$$n_{-\sigma} B_{D;k}^{-\sigma} + n_{-\sigma} B_S^{-\sigma} . \qquad (93)$$

The function $B_{D;k}^{-\sigma}$, called the bandwidth correction by the authors, depends on the wave vector **k** and is given by

$$B_{D;k}^{-\sigma} = \frac{1}{n_{-\sigma}(1-n_{-\sigma})} \frac{1}{N} \sum_{ij} (-t_{ij}) e^{-i\mathbf{k}\cdot(\mathbf{r}_i - \mathbf{r}_j)} \left( \langle \hat{n}_{i-\sigma} \hat{n}_{j-\sigma} \rangle - n_{-\sigma}^2 \right. \\ \left. - \langle c_{j\sigma}^+ c_{j-\sigma}^+ c_{i-\sigma} c_{i\sigma} \rangle - \langle c_{j\sigma}^+ c_{i-\sigma}^+ c_{j-\sigma} c_{i\sigma} \rangle \right) , \qquad (94)$$

where the three parts were interpreted as density correlation, double hopping, and spin exchange. The **k**-independent term $B_S^{-\sigma}$ is the bandshift correction and is given by

$$B_S^{-\sigma} = \frac{1}{n_{-\sigma}(1-n_{-\sigma})} \frac{1}{N} \sum_{ij} (-t_{ij}) \langle c_{i-\sigma}^+ c_{j-\sigma} (2\hat{n}_{i\sigma} - 1) \rangle . \qquad (95)$$

In the MAA method the authors[35] used the CPA equations [Eq. (55) above] with two centers of gravity $V_i^\sigma$ and probabilities $P_i^\sigma$ modified by the bandshift $B_S^{-\sigma}$ parameter. Contrary to the normal CPA results, the MAA method in the $U \gg D$ limit produced a self-consistent ferromagnetic solution in the middle of the lower Hubbard sub-band (for electron occupations $0.65 < n < 0.75$). This range of existence of spontaneous magnetization obtained at the strong Coulomb interaction was larger than the ranges obtained at smaller $U$.[35] In our model the range of ferromagnetism is also at the maximum for $U \gg D$. For smaller $U$ we have obtained a smaller range of ferromagnetism (see Fig. 4) shifted towards the half-filled band.

Different densities of states used in different models have also an influence on the range of ferromagnetism, but they do not decide about its appearance. The above quoted range of ferromagnetism was calculated[35] using the tight binding density of states for the bcc lattice. With this DOS they obtained a weak ferromagnetism. Using the tight binding fcc DOS the same group[34] has obtained (in the MAA method) a strong ferromagnetism within the whole range of electron occupations $0 < n \leq 1$. The semi-elliptic DOS used by us resembles rather the tight binding bcc DOS, although it is not peaked as strongly at the half-filled point. Combined with our model it gives ferromagnetism for $0.32 < n < 0.87$.

In their numerical MAA calculations Nolting and co-workers[34,35] neglected the bandwidth correction. As opposed to this the bandwidth correction was included in our numerical calculations (see Figs. 1 and 3).

The correlation function given by Eq.(93) corresponds in our model and in our notation to the function

$$B_{k,\sigma}^T(\varepsilon) + S_\sigma^T(\varepsilon) \quad , \tag{96}$$

which was derived above rigorously in the Hubbard III approximation with added inter-site correlation. Only after making in our formula for the bandshift, Eq. (45), and the bandwidth, Eq. (48), a very simplistic assumption

$$F_H^\sigma(\varepsilon) C_- [\varepsilon - \Omega_\sigma^T(\varepsilon)] F_{H,0}^\sigma(\varepsilon) C_-(\varepsilon) \approx \frac{1}{(1 - n_{-\sigma})} \quad , \tag{97}$$

we obtain from our Eq. (96) the correlation function of Eq. (93) given by the SDA and MAA methods.

Summarizing the comparison with existing models: the MAA formulas have come out as a simplified version of the formulas derived analytically in this paper within the Hubbard III scheme which included the inter-site correlations: $\langle c_{i-\sigma}^+ c_{j-\sigma} \rangle$ and $\langle \hat{n}_{i\sigma} c_{i-\sigma}^+ c_{j-\sigma} \rangle$.

REFERENCES

[1] J. Hubbard, Proc. Roy. Soc. A **276**, 238 (1963).

[2] J. Hubbard, Proc. R. Soc. A **277**, 237 (1964).

[3] J. Hubbard, Proc. Roy. Soc. A **281**, 401 (1964).

[4] E. H. Lieb and F. Y. Wu, Phys. Rev. Lett. **20**, 1445 (1968).

[5] P. Soven, Phys. Rev. **156**, 809 (1967).

[6] B. Velický, S. Kirkpatrick, and H. Ehrenreich, Phys. Rev. **175**, 747 (1968).

[7] S.E. Barnes, J. Phys. F: Met. Phys. **6**, 1375 (1976).

[8] P. Coleman, Phys. Rev. B **29**, 3035 (1984).

[9] A. Georges, G. Kotliar, W. Krauth, and M.J. Rozenberg, Rev. Mod. Phys. **68**, 13 (1996).

[10] M. Jarrell, Phys. Rev. Lett. **69**, 168 (1992).

[11] M.J. Rozenberg, X.Y. Zhang, and G. Kotliar, Phys. Rev. Lett. **69**, 1236 (1992).

[12] A. Georges and W. Krauth, Phys. Rev. Lett. **69**, 1240 (1992).

[13] J.E. Hirsch and F. Marsiglio, Phys. Rev. B **39**, 11515 (1989).

[14] F. Marsiglio and J.E. Hirsch, Phys. Rev. B **41**, 6435 (1990).

[15] J.C. Amadon and J.E. Hirsch, Phys. Rev. B **54**, 6364 (1996).

[16] J.E. Hirsch, Phys. Rev. B **59**, 6256 (1999).

[17] R. Micnas, J. Ranninger, and S. Robaszkiewicz, Rev. Mod. Phys. **62**, 113 (1990); Phys. Rev. B **39**, 11653 (1989).

[18] C. Castellani, M. Grilli, and G. Kotliar, Phys. Rev. B **43**, 8000 (1991).

[19] G. Górski and J. Mizia, Physica C **309**, 138 (1998).

[20] K. Penc and A. Zawadowski, Phys. Rev. B **50**, 10578 (1994)

[21] W.A. Atkinson and J.P. Carbotte, Phys. Rev. B **51**, 1161 (1995); **51**, 16371 (1995).

[22] V.J. Emery, Phys. Rev. Lett. **58**, 2794 (1987).

[23] P. B. Littlewood, C. M. Varma, and E. Abrahams, Phys. Rev. Lett. **63**, 2602 (1989).

[24] P. B. Littlewood, Phys. Rev. B **42**, 10075 (1990).

[25] Y. Bang, G. Kotliar, C. Castellani, M. Grilli, and R. Raimondi, Phys. Rev. B **43**, 13724 (1991).

[26] Y. Bang, G. Kotliar, R. Raimondi, C. Castellani, and M. Grilli, Phys. Rev. B **47**, 3323 (1993).

[27] R. Raimondi, C. Castellani, M. Grilli, Yunkyu Bang, and G. Kotliar, Phys. Rev. B **47**, 3331 (1993).

[28] H. Fukuyama and H. Ehrenreich, Phys. Rev. B **7**, 3266 (1973).

[29] J. Mizia, phys. stat. sol. (b) **74**, 461 (1976).

[30] L.M. Roth, Phys. Rev. **184**, 451 (1969).

[31] G. Geipel and W. Nolting, Phys. Rev. B **38**, 2608 (1988).

[32] W. Nolting and W. Borgiel, Phys. Rev. B **39**, 6962 (1989).

[33] T. Herrmann and W. Nolting, J. Magn. Mat. **170**, 253 (1997).

[34] M. Potthoff, T. Herrmann, T. Wegner, and W. Nolting, phys. stat. sol. (b) **210**, 199 (1998).

[35] T. Herrmann and W. Nolting, Phys. Rev. B **53**, 10579 (1996).

[36] D.N. Zubarev, Usp. Fiz. Nauk **71**, 71 (1960). Translation Sov. Phys. Usp. **3**, 320 (1960).

[37] G. Górski, J. Mizia, and K. Kucab, Physica B **336**, 308 (2003).

[38] G. Górski, J. Mizia, and K. Kucab, Physica B **325**, 106 (2003).

**FIGURE CAPTIONS**

FIG. 1 Dependence of critical on-site Stoner exchange field, $F_{in}^{cr}$, on the electron occupation $n$. Solid line –extended Hubbard model with high approximation for the bandwidth $B_{k,\sigma}^{T}$ and bandshift $S_{\sigma}^{T}(\varepsilon)$ factors, for $U=15D$. Dashed line –the H-F approximation for $B_{k,\sigma}^{T}$ and $S_{\sigma}^{T}$, the case of strong Coulomb correlation, $U/D \to \infty$. Dotted line – the CPA result, the case of strong Coulomb correlation, $U/D \to \infty$.

FIG. 2 Dependence of critical on-site Stoner exchange field, $F_{in}^{cr}$, on the electron occupation $n$ after including the bandshift factor $S_{\sigma}^{T}(\varepsilon)$ (the factor $B_{k,\sigma}^{T}=0$). Solid line – extended Hubbard model with high approximation for the bandshift factor $S_{\sigma}^{T}(\varepsilon)$, $U=15D$. Dashed line – the H-F approximation for $S_{\sigma}^{T}$, the case of strong Coulomb correlation, $U/D \to \infty$. Dotted line – the CPA result, the case of strong Coulomb correlation, $U/D \to \infty$.

FIG. 3 The dependence of critical on-site Stoner exchange field, $F_{in}^{cr}$, on the electron occupation $n$ after including the bandwidth factor $B_{k,\sigma}^{T}$ (the factor $S_{\sigma}^{T}(\varepsilon)=0$). Solid line – extended Hubbard model with high approximation for the bandwidth factor $B_{k,\sigma}^{T}$, $U=15D$. Dashed line – the H-F approximation for $B_{k,\sigma}^{T}$, the case of strong Coulomb correlation, $U/D \to \infty$. Dotted line – the CPA result, the case of strong Coulomb correlation, $U/D \to \infty$.

FIG. 4 The dependence of critical on-site Stoner exchange field, $F_{in}^{cr}$, on the electron occupation $n$ for different Coulomb interactions $U$. Both: the bandwidth and the bandshift factors are included. Solid line – $U=15D$, dashed line – $U=10D$, and dotted line – $U=5D$.

FIG. 5 The dependence of Curie temperature on the electron occupation $n$ for different Coulomb interactions $U$. Solid line – $U=15D$, dashed line – $U=10D$, and dotted line – $U=5D$. The $T_C$ scale on the right correspond to $D=1\text{eV}$.

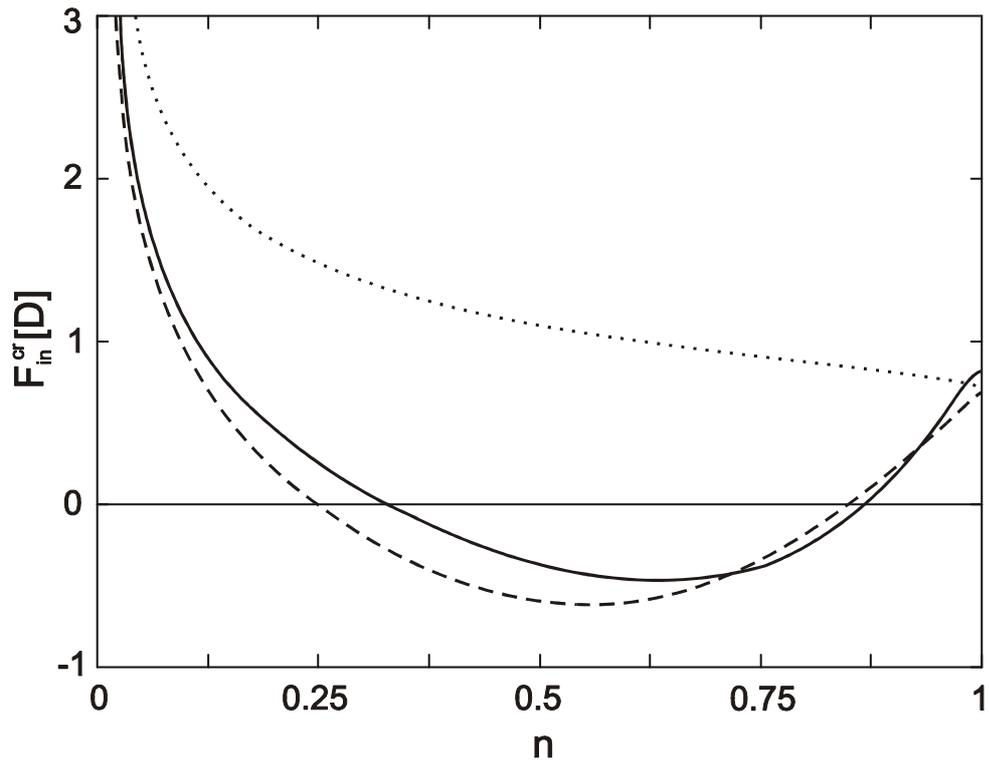

**FIG. 1 G. Górski**

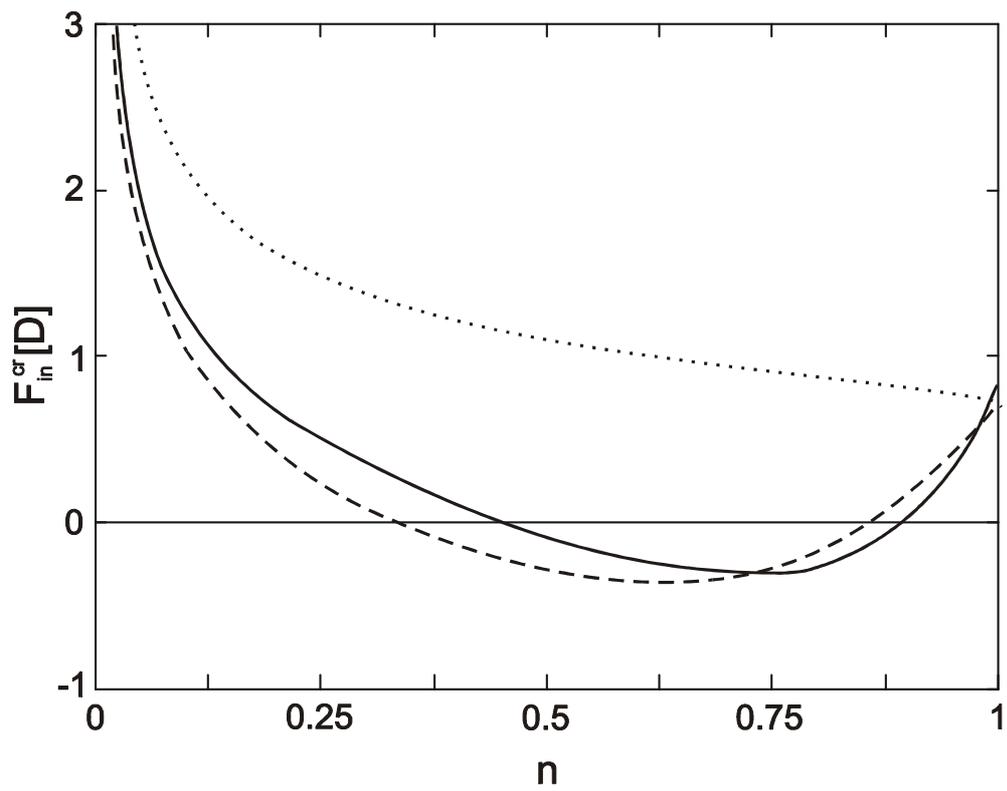

**FIG. 2** G.Górski

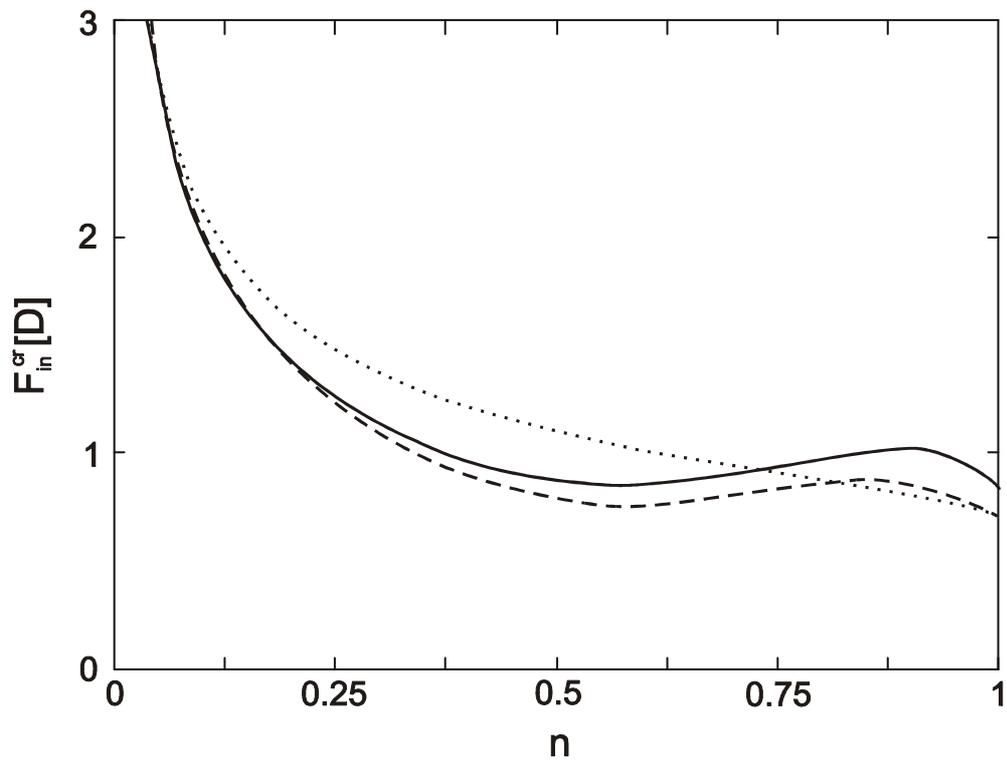

FIG. 3  G. Górski

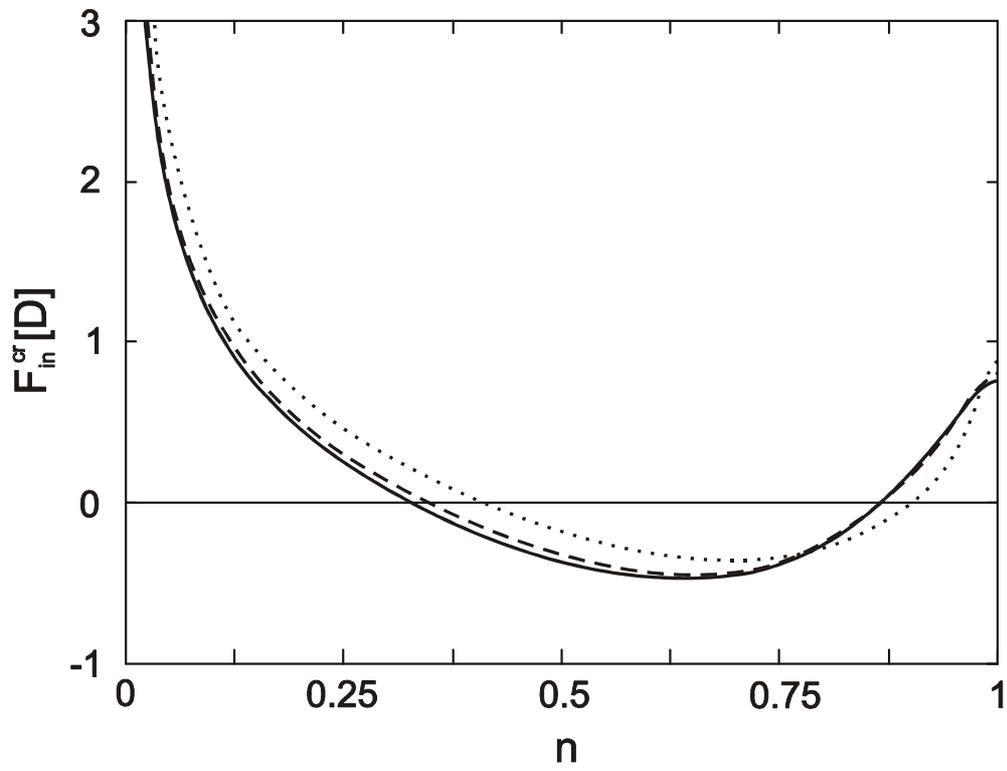

**FIG. 4** G. Górski

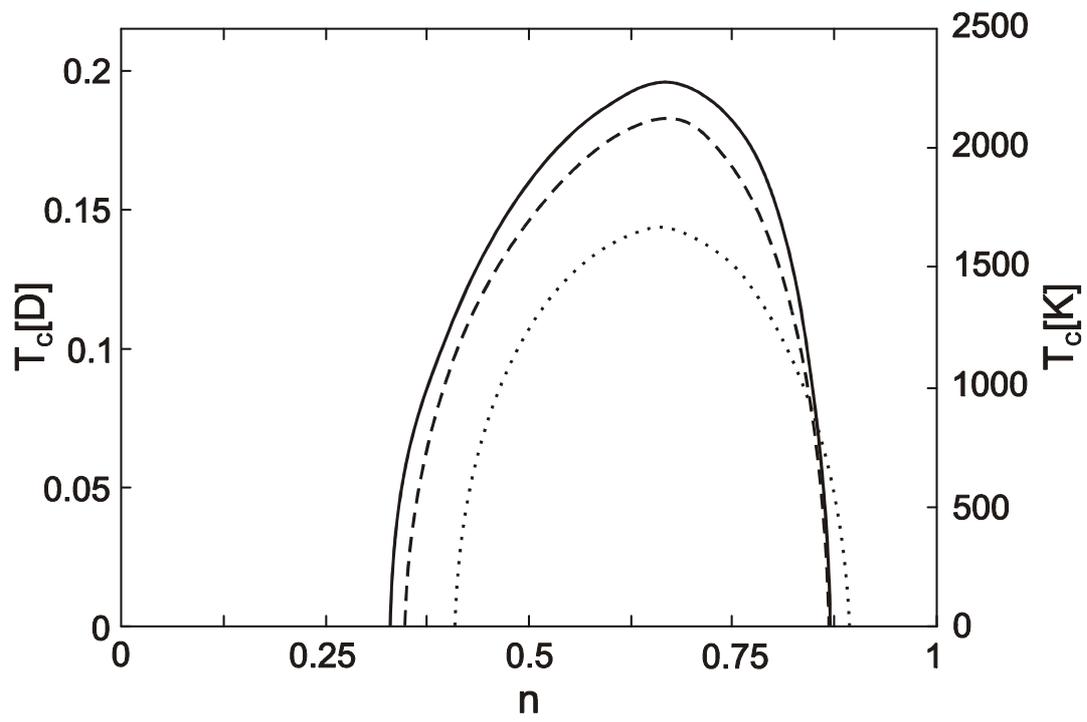

FIG. 5 G. Górski